\documentclass[longauth]{aa}

\usepackage[varg]{txfonts}
\usepackage{graphicx}
\usepackage{float}
\usepackage{natbib}
\bibpunct{(}{)}{;}{a}{}{,}

\usepackage{threeparttablex}
\usepackage{threeparttable}
\usepackage{longtable}
\usepackage{lscape}
\usepackage{bm}
\usepackage{verbatim}
\usepackage{makecell}
\usepackage{caption}
\usepackage{subcaption}
\usepackage{multirow}
\usepackage{booktabs}
\usepackage{colortbl}
\usepackage{xcolor}

\usepackage[colorlinks,linkcolor=blue,urlcolor=blue,citecolor=blue]{hyperref}
\begin{document}
\raggedbottom

\title{X-ray Characterization of Galaxy Groups and Hickson Groups in COSMOS-Web to $z\sim3.8$}
\subtitle{I. X-ray Luminosity Function Evolution}
\titlerunning{X-ray Characterization of COSMOS-Web Groups (Paper~I)}

\author{Ghassem Gozaliasl\inst{\ref{Aalto},\ref{Helsinki}}\thanks{e-mail: ghassem.gozaliasl@aalto.fi}
\and
G\"{u}nther Hasinger\inst{\ref{DZA},\ref{Dresden}}
\and
Gavin Leroy \inst{\ref{Durham}} 
\and
Alexis Finoguenov\inst{\ref{Helsinki}}
\and
Fatemeh Abedini\inst{\ref{Aalto},\ref{Helsinki},\ref{IASBS}}\and
 Hollis~B.~Akins\inst{\ref{UAT}}\and
Rafael~C.~Arango-Toro\inst{\ref{LAM}} \and
Arif Babul\inst{\ref{Victoria},\ref{Infosys}} \and
Caitlin M.~Casey\inst{\ref{ucsb}, \ref{UAT}, \ref{DAWN}} \and
Nicole E. Drakos \inst{\ref{HawaiiHilo}} \and
Andreas L. Faisst\inst{\ref{Pasadena}} \and
Carter Flayhart\inst{\ref{Rochester}} \and
Maximilien~Franco\inst{\ref{CEA},\ref{UAT}} \and
Zohreh Ghaffari\inst{\ref{INAF_trieste},\ref{IFPU}} \and
Santosh~Harish\inst{\ref{Rochester}} \and
Hossein Hatamnia\inst{\ref{Riverside}} \and
Olivier~Ilbert\inst{\ref{LAM}} \and
Jeyhan~S.~Kartaltepe\inst{\ref{Rochester}} \and
Ali Ahmad Khostovan\inst{\ref{Kentucky},\ref{Rochester}} \and
Maarit Korpi-Lagg\inst{\ref{Aalto}}\and
Ronaldo Laishram\inst{\ref{NAOJ}}\and
Daizhong Liu\inst{\ref{PurpleMObs}} \and
Matteo Maturi \inst{\ref{Heidelberg},\ref{ITP}} \and
Henry~Joy~McCracken\inst{\ref{IAP}} \and
Jed McKinney\inst{\ref{UAT}} \and
Nicolas McMahon\inst{\ref{Rochester}} \and
Lauro Moscardini\inst{\ref{Bologna},\ref{INAF},\ref{INFN}} \and
Mahdiyar Mousavi-Sadr\inst{\ref{ipm}}
\and
Jason Rhodes\inst{\ref{Pasadena}} \and
Brant E. Robertson\inst{\ref{Santacruz}}\and
Louise Paquereau\inst{\ref{IAP}} \and
Rasha M. Samir \inst{\ref{NRIAG}}\and 
Mark Sargent\inst{\ref{EPFL}} \and
Diana Scognamiglio\inst{\ref{JPL}} \and
Marko Shuntov\inst{\ref{DAWN},\ref{NBI},\ref{UG}} \and
Greta Toni\inst{\ref{Bologna},\ref{INAF},\ref{Heidelberg}}\and
Francesco Ubertosi\inst{\ref{INAF},\ref{Bologna}}\and
Eleni Vardoulaki\inst{\ref{TLS},\ref{eleni2}}}

\institute{Department of Computer Science, Aalto University, PO Box 15400, 00076, Espoo, Finland \label{Aalto}
\and
Department of Physics, University of Helsinki, Gustaf H\"{a}llstr\"{o}m katu 2, 00560 Helsinki,
Finland \label{Helsinki}
\and
Deutsches Zentrum f\"{u}r Astrophysik, Postplatz 1, 02826, G\"{o}rlitz, Germany \label{DZA}
\and
TU Dresden, Institute of Nuclear and Particle Physics, 01062, Dresden, Germany; DESY, Notkestrasse
85, 22607, Hamburg, Germany \label{Dresden}
\and
Institute for Computational Cosmology, Department of Physics, Durham University, South Road, Durham DH1 3LE, United Kingdom \label{Durham}
\and
Institute for Advanced Studies in Basic Sciences (IASBS)
444 Prof. Yousef Sobouti Blvd.,
Zanjan 45137-66731, Iran\label{IASBS}
\and
The University of Texas at Austin, 2515 Speedway Blvd Stop C1400, Austin, TX 78712, USA \label{UAT}
\and
Aix Marseille Univ, CNRS, LAM, Laboratoire d'Astrophysique de Marseille, Marseille, France \label{LAM}
\and
Department of Physics and Astronomy, University of Victoria, BC V8X 4M6, Canada \label{Victoria}
\and
Infosys Visiting Chair Professor, Indian Institute of Science, Bangalore 560012, India \label{Infosys}
\and
Department of Physics, University of California, Santa Barbara, Santa Barbara, CA 93106 USA \label{ucsb}
\and
Cosmic Dawn Center (DAWN), Denmark \label{DAWN}
\and
Department of Physics, University of Hawaii, Hilo, 200 W Kawili St, Hilo, HI 96720, USA \label{HawaiiHilo}
\and
Caltech/IPAC, MS 314-6, 1200 E. California Blvd. Pasadena, CA 91125, USA \label{Pasadena}
\and
Laboratory for Multiwavelength Astrophysics, School of Physics and Astronomy, Rochester Institute of Technology, 84 Lomb Memorial Drive, Rochester, NY 14623, USA \label{Rochester}
\and
Université Paris-Saclay, Université Paris Cité, CEA, CNRS, AIM, 91191 Gif-sur-Yvette, France \label{CEA}
\and
INAF-Osservatorio Astronomico di Trieste, via G.B. Tiepolo 11, 34143 Trieste, Italy \label{INAF_trieste}
\and
 IFPU: Institute for Fundamental Physics of the Universe, Via Beirut, 2, 34151, Italy \label{IFPU}
 \and
Department of Physics and Astronomy, University of California, Riverside, 900 University Avenue, Riverside, CA 92521, USA \label{Riverside}
\and
Department of Physics and Astronomy, University of Kentucky, 505 Rose Street, Lexington, KY 40506, USA \label{Kentucky}
\and
  NAOJ: National Astronomical Observatory of Japan, 2-21-1 Osawa, Mitaka, Tokyo 181-8588, Japan \label{NAOJ}
\and
Purple Mountain Observatory, Chinese Academy of Sciences, 10 Yuanhua Road, Nanjing 210023, China \label{PurpleMObs}
\and
Zentrum für Astronomie, Universität Heidelberg, Philosophenweg 12, 69120 Heidelberg, Germany \label{Heidelberg}
\and
ITP, Universität Heidelberg, Philosophenweg 16, 69120, Heidelberg, Germany \label{ITP}
\and
Institut d'Astrophysique de Paris, UMR 7095, CNRS, and Sorbonne Université, 98 bis boulevard Arago, 75014 Paris, France \label{IAP}
\and
University of Bologna - Department of Physics and Astronomy "Augusto Righi" (DIFA), Via Gobetti 93/2, I-40129 Bologna, Italy \label{Bologna}
\and
INAF - Osservatorio di Astrofisica e Scienza dello Spazio di Bologna, via Gobetti 93/3, 40129 Bologna, Italy \label{INAF}
\and
INFN – Sezione di Bologna, Viale Berti Pichat 6/2, 40127, Bologna, Italy \label{INFN}
\and
School of Astronomy, Institute for Research in Fundamental Sciences (IPM), P. O. Box 19395-5531, Tehran, Iran
  \label{ipm}
\and
 Department of Astronomy and Astrophysics, University of California, Santa Cruz, 1156 High Street, Santa Cruz, CA 95064, USA \label{Santacruz}
 \and
National Research Institute of Astronomy and Geophysics (NRIAG), Cairo, Egypt \label{NRIAG}
\and
EPFL Laboratory of Astrophysics (LASTRO), Observatoire de Sauverny, CH – 1290 Versoix, Switzerland \label{EPFL}
\and
Jet Propulsion Laboratory, California Institute of Technology, 4800 Oak Grove Drive, Pasadena, CA 91001, USA \label{JPL}
\and
Niels Bohr Institute, University of Copenhagen, Jagtvej 128, 2200 Copenhagen, Denmark \label{NBI}
\and
University of Geneva, 24 rue du Général-Dufour, 1211 Genève 4, Switzerland \label{UG}
\and
Thüringer Landessternwarte, Sternwarte 5, 07778, Tautenburg, Germany \label{TLS}
\and
IAASARS/National Observatory Athens, Hill of the Nymphs,
Athens, 11851, Greece \label{eleni2}
  }

 
\date{February 25, 2026}

\abstract{
Galaxy groups are the dominant halo environment in the Universe and a sensitive laboratory for
feedback and gas-cycling physics, yet a uniform, intermediate-redshift X-ray characterization of
the group population -- particularly compact groups -- has been limited by the availability of deep,
wide-area imaging needed to identify them. We present a uniform X-ray characterization of the
COSMOS-Web galaxy group population: 1,678 groups (CW-All) identified with the Adaptive Matched
Identifier of Clustered Objects (AMICO) algorithm applied to the JWST-selected COSMOS-Web
photometric catalog \citep{Toni2025}, over $0.45~\mathrm{deg}^2$ spanning $0.08 \lesssim z \lesssim
3.7$, and a parallel sample of 993 Hickson compact groups (CW-HCG; Hasinger et al. in prep.)
identified over the same sky area and redshift range ($0.08 \lesssim z \lesssim 3.82$). We
present a uniform aperture-photometry X-ray pipeline applied to the combined Chandra and XMM-Newton COSMOS
mosaic, measuring fluxes, luminosities, temperatures, and halo masses and radii for every group,
together with detections, upper limits, and quality flags. We
detect X-ray emission from 499 CW-All groups (29.7\%) and 332 CW-HCG groups (33.4\%) at SNR~$\geq
2.0$, and we present the X-ray luminosity function of the detected population in four redshift bins.
The characteristic luminosity $L^\star$ rises monotonically by $\sim$1.9--2.1~dex from the lowest to
the highest redshift bin in both catalogs, a trend that persists under two independent
cross-check binning schemes and is not explained by mass or richness selection of the detected
population, as traced by four independent, X-ray-independent proxies. A targeted search for diffuse
X-ray emission from the AmicOne protocluster candidate at $2.50 \lesssim z < 3.09$ finds no statistically significant signal at
the current depth. We publicly release the X-ray measurements and derived properties for the AMICO-selected and
compact-group catalogs, to serve as reference data products for this survey. Subsequent papers in this series—including studies of the weak-lensing halo-mass–$L_X$ scaling relation, X-ray stacking, dynamical masses, and galaxy environments—build on the X-ray measurements presented here.
}

\keywords{galaxies: clusters -- galaxies: groups -- X-rays: galaxies: clusters -- catalogs -- surveys}

\maketitle
\nolinenumbers

\section{Introduction}

Galaxy groups are the dominant halo environment of galaxies and the bridge between isolated galaxies
and rich clusters (see reviews by \citealp{Oppenheimer2021} and \citealp{Lovisari2021} and papers cited therein).
In this mass regime
($\sim10^{12}$--$10^{14}\,M_{\odot}$), non-gravitational heating and cooling can strongly reshape
the intragroup medium, making groups a sensitive laboratory for feedback physics and baryon cycling
\citep[e.g.][]{Babul2002,OSullivan2017,Hough2024,Popesso2024}. X-ray measurements of diffuse hot gas therefore provide direct constraints on
halo potentials, gas content, and assembly state \citep{Finoguenov2007,George2011}.

Recent wide-area X-ray studies illustrate both the progress and current limits of group
measurements. The SRG/eROSITA All-Sky Survey has delivered the largest current X-ray-selected
cluster and group catalog, with more than 12,000 systems to $z\approx1.3$ \citep{Bulbul2024}. In the
local Universe, eROSITA/eFEDS stacking has constrained average hot-gas
properties down to group and near Milky-Way scales and shown that scatter in the $L_{\rm X}$--$M$
relation is tied to structural variations in gas distribution at fixed halo mass
\citep{Popesso2024,Marini2024}, as anticipated by \citep{McCarthy2004,McCarthy2008}.
A substantial fraction of optically selected groups are X-ray faint
relative to detected systems of similar mass \citep{Pearson2017},
consistent with active galactic nucleus (AGN) feedback-regulated gas redistribution
and motivating homogeneous catalogs that retain both detections and upper limits.

At higher redshift, a complete X-ray census remains challenging: systematic surveys beyond
$z\gtrsim1.5$ are still sparse \citep{Willis2020,Toni2024}, and detections at very high redshift
($z \gtrsim 4$) are rare. A key milestone is the Chandra+JWST detection of extended X-ray emission from a proto-cluster
at $z\approx5.7$, showing that shock-heated intracluster gas can already be in place at
very early cosmic time \citep{Bogdan2026}. Together, low-$z$ stacking constraints and rare high-$z$
detections define the present frontier and underscore the need for uniform, intermediate-redshift
group catalogs with consistent X-ray measurements.

The COSMOS survey \citep{Scoville2007,Koekemoer2007,Taniguchi2007} is well suited to this step.
Previous XMM-Newton+Chandra work identified 247 X-ray groups at
$0.08\leq z<1.53$ \citep{Finoguenov2007,George2011,Smolcic2011,Gozaliasl2019}; COSMOS-Web extends group
selection with JWST depth and resolution over a wide contiguous area \citep{Casey2023}. The
AMICO-based catalog contains 1,678 systems over $0.45~\mathrm{deg}^2$ \citep{Toni2025}, with broad
redshift coverage and substantial spectroscopic support \citep{Shuntov2025,Khostovan2025}. In parallel, an
independently constructed Hickson-style compact-group sample -- a small number ($n_g\geq3$) of
galaxies in close angular proximity and small relative velocities \citep{Hickson1982} -- in the same
footprint provides a complementary high-density-environment selection (Hasinger et al. in prep.).

Compact groups are especially valuable because they isolate systems where tidal interactions,
accelerated galaxy transformation, and AGN triggering may be enhanced relative to looser
associations \citep{Hickson1982,MendesDeOliveira1994}. Their X-ray phenomenology is diverse, from
bright diffuse atmospheres to weak or undetected hot gas
\citep{Mulchaey2000,Osmond2004,Sun2009,Lovisari2015}. Measuring compact and general group samples
with the same X-ray data and pipeline enables a controlled comparison of selection effects and
physical demographics. The halo masses reported here adopt the weak-lensing calibrated $M$--$L_{\rm
X}$ relation of \citet{Leroy2026} (in prep.), who calibrate the group-scale $M_{200}$--$L_{\rm X}$ relation
directly in the COSMOS-Web footprint using deep JWST weak-lensing halo masses
\citep{Scognamiglio2026}, superseding the \citet{Leauthaud2010} relation used in earlier
COSMOS group studies.

In this paper, we present a standalone reference release of COSMOS-Web group X-ray measurements,
using a
uniform aperture-photometry X-ray pipeline applied to all systems in both samples. We report
detections and upper limits, derive fluxes, luminosities, temperatures, and halo-mass proxies, and
present the X-ray luminosity function of detected systems as the main statistical result. The
emphasis is on reproducible X-ray measurement and calibrated data products; detailed physical
interpretation (e.g., AGN demographics, scaling-relation physics, and environmental trends) is
deferred to dedicated follow-up studies.

The paper is organized as follows. Sect.~\ref{sec:data} presents the imaging, spectroscopic, and
X-ray datasets; Sect.~\ref{sec:groupcat} describes the group catalogs, membership, and brightest
group galaxy (BGG) identification; Sect.~\ref{sec:analysis} details the X-ray measurement pipeline;
Sect.~\ref{sec:catalog_overview} summarizes the released catalog properties; Sect.~\ref{sec:xlf}
presents the X-ray luminosity function analysis; Appendix~\ref{app:release} documents the public data
products; and Sect.~\ref{sec:conclusions} summarizes the main results. Throughout this paper, we
adopt a flat $\Lambda$CDM cosmology consistent with Planck 2018 \citep{Aghanim2020}, with $H_0 =
67.4\,\mathrm{km\,s^{-1}\,Mpc^{-1}}$ and $\Omega_{\rm m}=0.315$.

\section{Data}
\label{sec:data}

\subsection{COSMOS-Web imaging and photometric/spectroscopic catalogs}

COSMOS-Web provides deep JWST imaging over the COSMOS field, covering $0.54~\mathrm{deg}^2$ with
NIRCam (F115W, F150W, F277W, F444W) and $0.19~\mathrm{deg}^2$ with MIRI/F770W \citep{Casey2023}. The
NIRCam data reach typical $5\sigma$ depths of $26.6$--$27.3$ (F115W), $26.9$--$27.7$ (F150W),
$27.5$--$28.2$ (F277W), and $27.5$--$28.2$ mag (F444W) in $0\farcs15$ apertures.

For this work, we use the COSMOS-Web photometric catalog products (photometric redshifts, stellar
masses, and structural parameters) together with the matched COSMOS spectroscopic compilation for
redshift validation and group-member confirmation \citep{Shuntov2025,Khostovan2025}. This combined
dataset is the input sample for AMICO group detection \citep{Toni2025} and the downstream membership and X-ray
analyses presented in this paper (Sects.~\ref{sec:groupcat} and \ref{sec:analysis}).

\subsection{X-ray data: Chandra and XMM-Newton}
\label{sec:xraydata}

The X-ray data used in this work are the combined Chandra and XMM-Newton observations of the COSMOS
field, following the methodology described in \citet{Gozaliasl2019}. The XMM-Newton Wide-Field
Survey in the COSMOS field \citep{Hasinger2007} provides deep coverage of the full
$2~\mathrm{deg}^2$ COSMOS field, with 55 pointings totaling approximately 1.5~Ms of exposure and an
average vignetting-corrected depth of 40~ks. The Chandra COSMOS-Legacy Survey
\citep{Elvis2009,Civano2016,Marchesi2016} provides high-resolution imaging that improves X-ray
centroid precision from 15$''$ (XMM-only) to 5$''$ \citep{Gozaliasl2019}, which is important for
measuring BGG--X-ray-peak offsets. The combined map reaches a flux limit of $3 \times
10^{-16}~\mathrm{erg~cm^{-2}~s^{-1}}$ in the $0.5$--$2.0$~keV band. Maps are distributed with $4''$
pixels, conservative point-source masking, and homogenized exposure weighting, together with
companion $1\sigma$ uncertainty images, in two versions: a full map, and a masked map with all
detected X-ray sources (out to $2\,R_{200}$ of each detected group) as well as point sources
excised, leaving only clean, diffuse background emission. We use the full map for background
determination and X-ray characterization of groups at $z<1.5$, and the masked map at $z\geq1.5$, to
minimize contamination from the emission of projected, unrelated foreground and background
structures, which becomes increasingly important as the projected source density on the sky rises
with redshift; the masked map is used throughout for the diffuse, stacking-oriented analyses of this
series (Paper~II). Fig.~\ref{fig:xray_map_masked} shows the full-field masked map.

\begin{figure*}[htbp]
\centering
\includegraphics[width=0.72\textwidth]{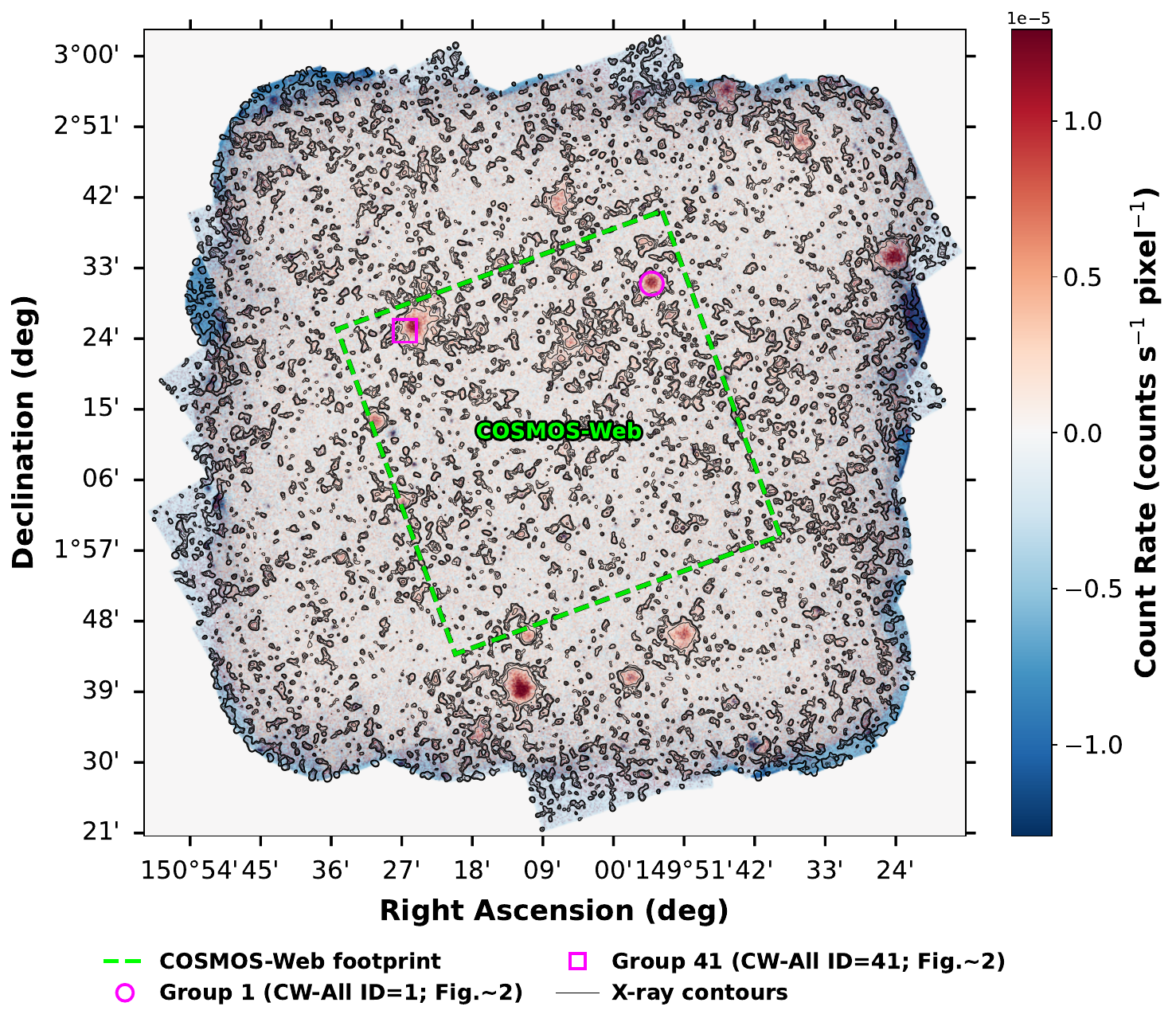}
\caption{Full COSMOS X-ray map in the 0.5--2.0~keV band, combining Chandra and XMM-Newton
observations. The map covers approximately 2~deg$^2$ and shows background-subtracted count rates
with black X-ray contours overlaid, drawn at the 60th, 71.7th, 83.3rd, and 95th percentiles of the
Gaussian-smoothed ($\sigma \approx 3.2$~pix), positive-valued count-rate distribution. The lime
dashed polygon indicates the COSMOS-Web footprint
(0.54~deg$^2$). The magenta circle marks Group~1
($\mathrm{RA}=149.919\degr$, $\mathrm{Dec}=2.517\degr$; SNR=28.2, $z =
0.710$) and the magenta square marks Group~41
($\mathrm{RA}=150.444\degr$, $\mathrm{Dec}=2.417\degr$; SNR=39.3, $z = 0.155$), both shown zoomed
in Fig.~\ref{fig:group1_showcase}.}
\label{fig:xray_map}
\end{figure*}

\begin{figure*}[htbp]
\centering
\begin{subfigure}{0.47\textwidth}
\centering
\includegraphics[width=\linewidth]{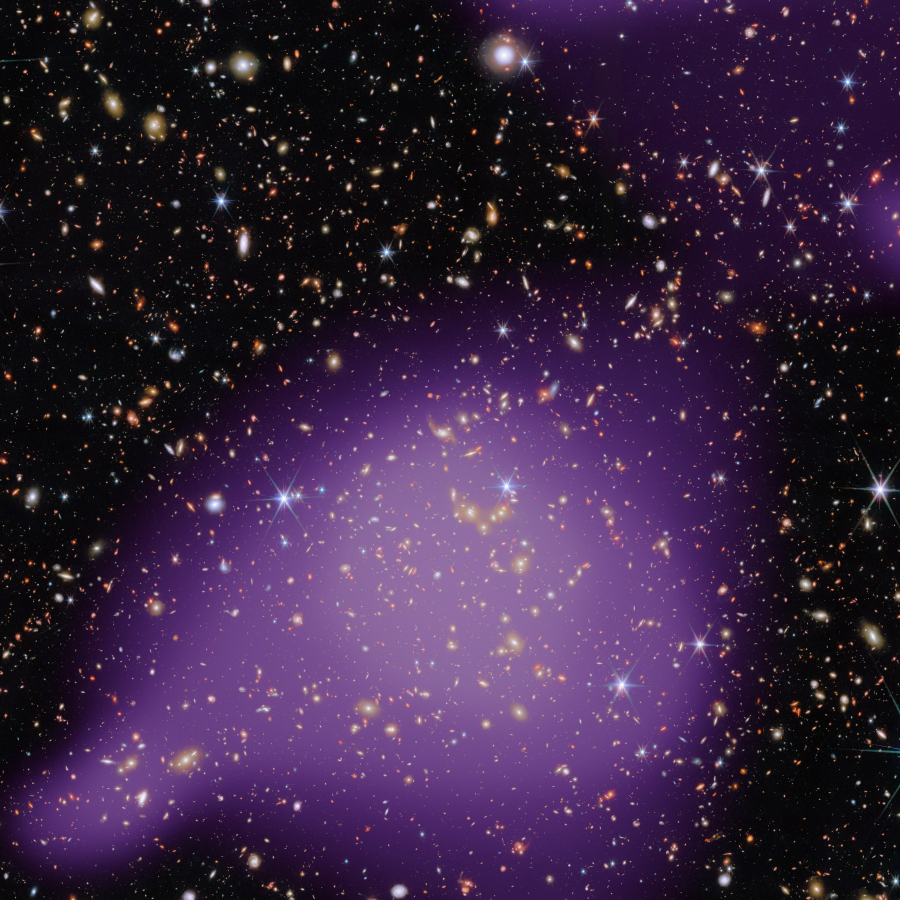}
\caption{}
\label{fig:group1_showcase_rgb}
\end{subfigure}
\hfill
\begin{subfigure}{0.47\textwidth}
\centering
\includegraphics[width=\linewidth]{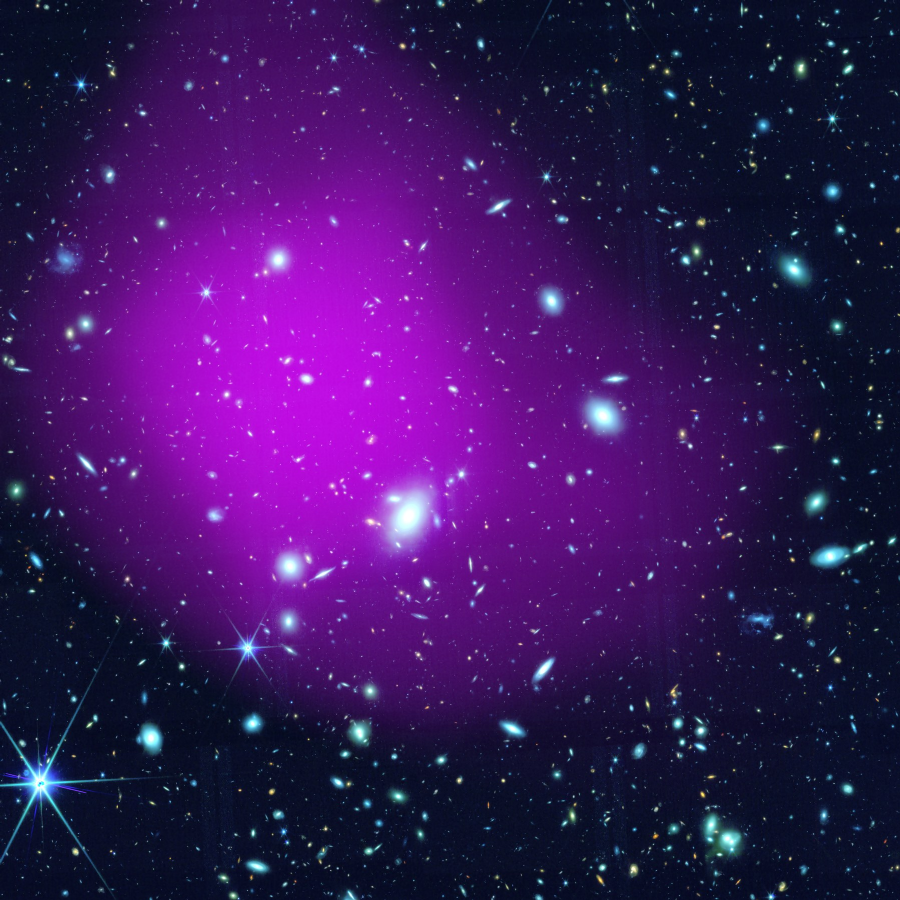}
\caption{}
\label{fig:group41_showcase_rgb}
\end{subfigure}
\caption{JWST/COSMOS-Web color composites with the diffuse X-ray emission overlaid (magenta/white
contours), showing member galaxies against the intragroup medium, for
\emph{(a)} Group~1 ($\mathrm{RA}=149.919\degr$, $\mathrm{Dec}=2.517\degr$;
SNR=28.2, $z=0.710$, the third-highest-significance detection in our sample,
Table~\ref{tab:xray_sample_all}, and its most massive, richly populated cluster-scale system,
$\lambda_\star = 77.6$, $T_{\rm X} = 4.0$~keV, $\log_{10}(L_{\rm X}/\mathrm{erg~s}^{-1}) = 43.7$,
$\log_{10}(M_{200}/M_\odot) = 14.4$) and
\emph{(b)} Group~41 ($\mathrm{RA}=150.444\degr$, $\mathrm{Dec}=2.417\degr$; SNR=39.3, $z=0.155$,
the highest-significance detection in our sample, $\lambda_\star = 3.6$, $T_{\rm X} = 0.6$~keV,
$\log_{10}(L_{\rm X}/\mathrm{erg~s}^{-1}) = 42.3$, $\log_{10}(M_{200}/M_\odot) = 13.6$).
Panel (a) credit: ESA/Webb, NASA \& CSA, G.\ Gozaliasl, A.\ Koekemoer, M.\ Franco;
\url{https://esawebb.org/images/comparisons/potm2504/}.}
\label{fig:group1_showcase}
\end{figure*}

Fig.~\ref{fig:xray_map} shows the full COSMOS X-ray map with the COSMOS-Web footprint overlaid,
marking the position of Group~1 (CW-All ID=1, SNR=28.2, $z=0.710$), the third-highest-significance
detection in our sample (Table~\ref{tab:xray_sample_all}) and its most massive, richly populated
cluster-scale system ($\lambda_\star = 77.6$, $T_{\rm X} = 4.0$~keV,
$\log_{10}(L_{\rm X}/\mathrm{erg~s}^{-1}) = 43.7$, $\log_{10}(M_{200}/M_\odot) = 14.4$, $R_{200} =
997$~kpc), and of Group~41 (CW-All ID=41, SNR=39.3, $z=0.155$), the highest-significance detection
in our sample ($\lambda_\star = 3.6$, $T_{\rm X} = 0.6$~keV,
$\log_{10}(L_{\rm X}/\mathrm{erg~s}^{-1}) = 42.3$, $\log_{10}(M_{200}/M_\odot) = 13.6$, $R_{200} =
603$~kpc). Fig.~\ref{fig:group1_showcase}
shows zoomed JWST/COSMOS-Web color composites with the X-ray emission overlaid for both
systems.

\section{Galaxy Group Catalogs}
\label{sec:groupcat}

\subsection{COSMOS-Web Group Catalog (Toni et al.\ 2025): parent group detection}

The parent group catalog, presented in \citet{Toni2025}, is built with the Adaptive Matched
Identifier of Clustered Objects (AMICO) algorithm \citep{Bellagamba2018,Maturi2019}. AMICO implements
an optimal matched-filter algorithm that assigns an optimal weight to each galaxy based on its
position, photometric redshift, and magnitude to produce unbiased three-dimensional amplitude maps
with minimum variance, and detects groups as peaks in this map. For
each detection, AMICO assigns galaxy membership probabilities, from which the apparent richness
$\lambda$ and intrinsic richness $\lambda_\star$ are computed (summing member probabilities above
the characteristic magnitude $m_\star + 1.5$, where $m_\star$ is evaluated per group from the
redshift-evolving Schechter luminosity function template below, not a single fixed value) within
$R_{200}$, the radius enclosing a mean density
200 times the critical density of the Universe at the group's redshift, within which the enclosed
mass is $M_{200}$; $R_{500}$ and $M_{500}$ are defined analogously for an overdensity of 500. The
membership probabilities are also
used in an iterative deblending scheme. For COSMOS-Web, AMICO was run with
amplitude-map resolution $\Delta r = 0.3'$, $\Delta z = 0.01$, using the JWST F150W magnitude as the
single galaxy property, a truncated NFW \citep{Navarro1997} cluster profile, and a redshift-evolving
Schechter luminosity function \citep{Toni2025}.

Applying AMICO to the COSMOS-Web galaxy catalog yields 1,678 groups over an effective area of
$0.45~\mathrm{deg}^2$ at $\mathrm{S/N_{nocl}} > 6.0$, extending to $z = 3.7$. \citet{Toni2025}
estimate a purity of $\sim$77\% for the full sample ($\sim$90\% for the $\sim$670 highest-S/N
systems) using data-driven mock catalogs generated with the Selection Function extrActor
(SinFoniA), in which AMICO is applied to synthetic galaxy catalogs built from the real
COSMOS-Web data with a known ``truth table'' of input groups. More
than 500 systems have spectroscopic confirmation \citep{Khostovan2025}. AMICO imposes no upper
mass or richness cut: the released catalog's intrinsic richness spans $\lambda_\star = 2.0$ (the
lower selection floor) to $\lambda_\star = 77.6$, extending into the cluster-richness regime rather
than being restricted to group-scale systems alone. We name this sample CW-All,
rather than CW-AMICO, because it is defined by population -- the full AMICO-selected group
catalog -- not by the group-finding algorithm alone; we use this name throughout this paper and its
data release.

\subsection{COSMOS-Web Hickson Compact Groups (Hasinger et al., in preparation)}
\label{sec:hcg_hasinger}

Within the same COSMOS-Web footprint, 993 Hickson compact groups (CW-HCG) are independently
identified by Hasinger et al.\ (in preparation), adapting the original \citet{Hickson1982}
criteria (high angular surface density and isolation from comparable-brightness neighbors) with a
relaxed minimum discovery membership of $n_g \geq 3$ member galaxies within 3~mag of the brightest
member, rather than the classic Hickson (1982) $\geq 4$. CW-HCG spans $0.08 \lesssim z \lesssim
3.82$ (median $z \approx 1.13$) and is selected independently of AMICO, so the two catalogs are not
mutually exclusive subsets but complementary group-finding methods applied to the same underlying
galaxy population. We release CW-HCG as a separate, parallel catalog rather than merging it into
CW-All.

\subsection{Group spectroscopic membership and BGG identification}
\label{sec:membership}

Group spectroscopic membership is determined with a probabilistic $P_z \times P_v$ model that assigns each
candidate galaxy a joint probability based on its photometric or spectroscopic redshift
consistency with the group redshift ($P_z$) and a Gaussian probability in its projected velocity
offset from the AMICO-assigned group center ($P_v$). The assignment is iterated to convergence on the group's systemic redshift and
line-of-sight velocity dispersion $\sigma_v$. The velocity dispersion is estimated using the
gapper method \citep{Beers1990}, which is robust at the modest spectroscopic richness typical
of this survey. Dynamical halo masses are derived from $\sigma_v$ using the
\citet{Munari2013} $\sigma_v$--$M_{200}$ scaling relation, providing a mass estimate
independent of the X-ray analysis.

The search apertures for the two catalogs are chosen to reflect their physical character.
For CW-All, the aperture follows an adaptive hierarchy: X-ray-detected groups use $R_{200}$
as determined from the X-ray luminosity via the \citet{Leroy2026} (in prep.) scaling relation
(Sect.~\ref{sec:scaling}), with a 250~kpc floor; groups with a dynamical mass estimate
but no X-ray detection use $R_{200}^{\rm dyn}$; and the remaining groups, which lack
both X-ray and dynamical constraints, are assigned a fixed 750~kpc aperture, comparable to the
group-scale search radii adopted for spectroscopic membership elsewhere in this series
\citep{Ghaffari2026}. For CW-HCG,
the compact-group geometry motivates smaller search apertures: X-ray-detected compact groups
use $R_{200}$ from the X-ray scaling relation with a 250~kpc floor, identical to CW-All,
but non-detected groups are assigned a fixed 350~kpc aperture rather than 750~kpc,
consistent with the systematically smaller physical scale of Hickson-type systems. The
BGG recentering tolerance is also tighter for CW-HCG (10~kpc convergence criterion versus
20~kpc for CW-All), reflecting the more compact configuration of member galaxies.

These choices were validated through an extensive internal exercise that found membership and
aperture tuning to be a sub-dominant source of uncertainty relative to intrinsic halo dynamical
scatter and finite-$N$ velocity-dispersion noise at the richness of this survey; individual-group
dynamical masses are consequently treated internally with a per-group confidence indicator rather
than as precision mass estimates; dynamical-mass data products themselves are not part of the
public release described here (Appendix~\ref{app:release}) and are deferred to a future companion
release, while group membership and richness are part of the parent group catalogs
\citep[][Hasinger et al., in preparation, for CW-HCG]{Toni2025}.

The BGG is identified for every group using a mass-floored hybrid rank
score that combines JWST-based stellar mass and rest-frame $R$-band luminosity, with a soft
preference for centrality, consistent with the hybrid mass/luminosity BGG selection approach applied
to COSMOS-Web groups in \citet{Gozaliasl2025}. For CW-All the centrality window is 300~kpc; for CW-HCG, reflecting
the physically more concentrated structure of compact groups, it is tightened to 100~kpc. In
both catalogs the BGG selection is iterated jointly with the membership step: after an initial
membership pass, the BGG is identified, the group center is shifted to the BGG position if the
offset exceeds the convergence tolerance, and the membership is recomputed from the new center
(up to three recentering iterations). BGG identity is additionally stability-tested by
independently re-scoring each group under three alternative aperture definitions; the resulting
agreement fraction is stored as a quality flag (\textit{BGG stability}).
\subsection{Public X-ray catalog and quality flags}

The catalog released with this work (Appendix~\ref{app:release}) presents the X-ray measurements and derived properties of the CW-All and CW-HCG group candidates. For each system, it includes the group identifier, sky position, redshift, X-ray detection significance, flux, luminosity, temperature, inferred halo mass, and characteristic radii. The group positions and redshifts are taken from the parent catalog and provide the reference information required for the X-ray analysis.

The ancillary multiwavelength catalogs are not merged into the released catalog as independent science products. Instead, cross-matches with the VLA-COSMOS 3~GHz catalog \citep{Smolcic2017} and the Chandra COSMOS-Legacy point-source catalog \citep{Civano2016,Marchesi2016}, together with inspection of the X-ray emission and nearby sources, are used to assess the reliability of each group detection and to assign quality flags.

The released catalog contains two principal flags. The \texttt{C} flag identifies systems whose measured X-ray emission may be affected by projected contamination, whereas the \texttt{F} flag identifies detections suspected of being false positives. Flagged systems are retained so that users can apply quality selections appropriate to their scientific analysis. The absence of a flag indicates that no significant problem was identified through the adopted checks, but does not guarantee that the measurement is free of all systematic uncertainty. The complete column definitions and descriptions of the quality flags are provided in Appendix~\ref{app:release}.

\subsection{CW-All and CW-HCG: two complementary group samples}

We release the X-ray measurements for CW-All and CW-HCG as two separate catalogs rather than a
single merged table. The two samples are constructed in their respective parent papers with
independent group-finding algorithms (AMICO matched-filter
versus Hickson angular-surface-density and isolation criteria), do not share a common group
identifier namespace, and differ in their membership apertures (750~kpc flexible / $R_{200}$
for CW-All; 350~kpc fixed / $R_{200}$ for CW-HCG), BGG centrality windows (300~kpc versus
100~kpc), and physical character: CW-All is the general-purpose, richness-complete group
catalog spanning a wide halo-mass range, while CW-HCG targets the high-density compact-group
regime well suited to studies of tidal interactions, accelerated galaxy transformation, and
potential AGN triggering. The two catalogs are not mutually exclusive: a galaxy or even a
group centre can appear in both if it satisfies both selection criteria independently. A
positional cross-match ($<30\arcsec$, with consistent redshift) finds a CW-All counterpart for 222
of the 993 CW-HCG groups (22.4\%), corresponding to 214 of the 1,678 CW-All groups (12.8\%); the
majority of CW-HCG systems therefore probe density peaks not independently identified by AMICO.
The match rate rises steeply with CW-All richness, from 12.8\% for the full sample to 25.0\% for
groups with $\lambda_\star \geq 20$, indicating that the two catalogs converge on the same
high-richness systems while CW-HCG additionally recovers a population of lower-richness,
high-density structures that fall below the AMICO detection threshold.
Table~\ref{tab:catalog_summary} summarizes the principal statistical properties of both
catalogs; both receive the identical X-ray pipeline described in Sect.~\ref{sec:analysis}.

\begin{table*}
\caption{Summary comparison of the two released group catalogs. X-ray detection fractions and median halo masses refer
  to the detected (SNR~$\geq 2.0$) subsample; see Table~\ref{tab:detection} for the full detection statistics.}
\label{tab:catalog_summary}
\centering
\begin{tabular}{lcc}
\hline\hline
 & CW-All & CW-HCG \\
\hline
Group-finder & AMICO & Hickson criteria \\
$N_{\rm groups}$ & 1678 & 993 \\
Redshift range & $0.08$--$3.70$ & $0.08$--$3.82$ \\
Median redshift & 1.10 & 1.13 \\
$N$ X-ray detected (\%) & 499 (29.7\%) & 332 (33.4\%) \\
Median discovery richness & $\lambda_\star = 7.2$ & $n_g = 3$ (71\% of sample; $n_g \geq 3$ by selection) \\
Median $M_{200}^{L_{\rm X}}$ (detected) & $3.0\times10^{13}~M_\odot$ & $2.7\times10^{13}~M_\odot$ \\
Median $R_{200}^{L_{\rm X}}$ (detected) & $381$~kpc & $408$~kpc \\
\hline
\end{tabular}
\end{table*}

\section{X-ray Analysis Pipeline}
\label{sec:analysis}

Every group in both catalogs is processed through the same uniform sequence of observational
steps, so that differences between CW-All and CW-HCG are not an artifact of the analysis choices
themselves; any remaining differences instead reflect a combination of astrophysics and of the two
group-finders' distinct selection functions and group types (AMICO's richness/matched-filter
selection versus CW-HCG's compact, high-density-contrast criteria; Sect.~\ref{sec:groupcat}), which
this uniform pipeline lets us compare on equal observational footing without disentangling further.
The pipeline uses adaptive apertures, a redshift-binned median background
technique, adaptive sigma clipping in the background annulus, and a two-pass scheme: an initial pass
with a fixed physical aperture for detection, followed by refinement of detected sources using true
$R_{500}$ and $R_{200}$ from scaling relations, adaptive background annuli, and X-ray peak
centering.

\noindent\textbf{Observables and conventions.} Photometry is performed on the observed 0.5--2.0~keV
band. Following \citet{Finoguenov2007}, we calculate the final X-ray luminosity from this
observed-band flux by applying the $K$-correction $K(z,T)$ (Sect.~\ref{sec:flux}), yielding the
rest-frame 0.5--2.0~keV luminosity $L_{\rm X}$; this is the primary observable reported for each
group. For sources above the detection threshold we report the measured $L_{\rm X}$, for the
remainder we report 95\% upper limits. We also derive intragroup medium temperature $T$, halo mass,
and characteristic radii $R_{200}$ and $R_{500}$ from scaling relations tied to $L_{\rm X}$.

\subsection{Spectral model, flux conversion, and K-correction}
\label{sec:flux}

Net count rates from the exposure-corrected mosaic are converted to observed-frame (0.5--2~keV)
energy flux using the count-rate-to-flux conversion factor of \citet{Finoguenov2007},
$C_{\rm XR} = 5.41 \times 10^{-12}~\mathrm{erg~cm^{-2}~count^{-1}}$, derived from an APEC
\citep{Smith2001} thermal-plasma spectrum with $kT = 2$~keV, $1/3$ solar abundance, and $z=0.2$,
consistent with the flux calibration inherited by the revised COSMOS X-ray group catalog of
\citet{Gozaliasl2019}. The mosaic is delivered in count-rate units (not raw counts), and we apply
this factor directly to the net count rate measured from it, without any additional
multiplicative correction for the number of contributing cameras. The observed-frame flux,
$F_{\rm X} = C_{\rm net}\,C_{\rm XR}$, is then converted to a rest-frame 0.5--2.0~keV luminosity
via $L_{\rm X} = 4\pi\, D_{\rm L}^2\, F_{\rm X}\,
K(z, T)$, where the $K$-correction is computed analytically from the same APEC spectral family
evaluated at $kT = 1$~keV; this is the only route from the observed band to the rest-frame band, as
the counts themselves are always measured in the observed 0.5--2~keV band. Angular-diameter and
luminosity distances use a Planck-2018 flat $\Lambda$CDM cosmology \citep{Aghanim2020} ($H_0 =
67.4$~km~s$^{-1}$~Mpc$^{-1}$, $\Omega_{\rm m} = 0.315$).

\subsection{Aperture definition, background, and detection}
\label{sec:aperture}

The baseline source aperture is a physical radius of 300~kpc, converted to angular size at each
group's redshift and clipped to $5''$--$120''$. First pass: photometry with the 300~kpc aperture and
a fixed 450--750~kpc background annulus for all groups, giving net counts, SNR, fluxes, and mass
proxies. Detection requires SNR~$\geq 2.0$ (Sect.~\ref{sec:detection}). Second pass (detections
only): the source aperture is refined to $R_{500}$ (or $R_{200}$ if unavailable) from the scaling
relations of Sect.~\ref{sec:scaling}, the background annulus becomes adaptive ($2$--$3\,R_{500}$), and the measurement is
recentered on the X-ray peak (flux-weighted centroid within $R_{500}$/$R_{200}$). Non-detections
retain the first-pass aperture and background. Fig.~\ref{fig:group1_snr_appendix} illustrates this
geometry.

Background is estimated from the source's own annulus after (i) masking pixels within any other
group's source aperture, to avoid bias from projected neighbors, and (ii) adaptive sigma clipping
(thresholds of $2.5\sigma$, $2.0\sigma$, $1.5\sigma$ are tried in sequence, retaining the least
aggressive threshold for which $\geq$5\% of annulus pixels survive). As a cross-check we
additionally compute a redshift-binned median background (sliding window $z \pm 0.05$, minimum 5
groups per window, expanded to $2\,\Delta z$ if needed); Appendix~\ref{app:background} quantifies
the agreement between the two schemes. The primary fluxes and luminosities reported here use the
annulus-based background.

Net counts are $C_{\rm net} = C_{\rm src} - C_{\rm bkg,scaled}$ with uncertainty $\sigma_{\rm net} =
\sqrt{\sigma_{\rm src}^2 + \sigma_{\rm bkg,scaled}^2}$; sources with unmasked aperture coverage
below 50\% are flagged and excluded from flux and mass estimates. Detection significance is
$\mathrm{SNR} = C_{\rm net}/\sigma_{\rm net}$.

\subsection{Detection significance threshold and upper limits}
\label{sec:detection}

We tested detection thresholds of SNR~$\geq 1.5$, $2.0$, $2.5$, and $3.0$ using a dedicated sweep
over the full pipeline; the resulting catalogs show qualitatively identical trends, with higher
thresholds yielding progressively fewer, brighter detections. We adopt SNR~$\geq 2.0$ as the
baseline (single-trial Gaussian false-alarm probability of 2.3\%), balancing robustness against
false positives with sufficient statistics for catalog-level and stacking comparisons.

For non-detections we report a 95\% confidence Gaussian upper limit,
\begin{equation}
C_{\rm UL} = \max\left(0,\ C_{\rm net} + 1.645\,\sigma_{\rm net}\right),
\label{eq:UL}
\end{equation}
i.e.\ the net count rate plus 1.645 Gaussian standard deviations, distinct from the single-trial
Gaussian false-alarm probability used above to set the SNR~$\geq2.0$ detection threshold, and
using the full net (source+background) propagated uncertainty $\sigma_{\rm net}$ rather than the
background uncertainty alone, since $\sigma_{\rm bkg}$ can be substantially smaller than
$\sigma_{\rm net}$ and would understate the true upper limit. As a quality-control check we flag
non-detections whose upper-limit flux falls below the first quartile of the detected flux
distribution; the first-quartile cut is an empirical choice, not a physically derived threshold, and
identifies the subset of upper limits most discrepant with the bulk of genuine detections. These
flagged systems are better described as insecure or unreliable non-detections -- most plausibly
associated with heavily masked apertures or low-coverage pointings -- rather than confirmed evidence
of genuinely faint intragroup emission; this diagnostic is presented in Sect.~\ref{sec:catalog_overview} together
with the other descriptive catalog statistics.

Groups below the individual detection threshold can additionally be characterized in an ensemble
sense via X-ray stacking in redshift bins. Stacking methodology, validation (PSF modeling,
random-position null tests), and results are presented in full in Paper~II, which applies the
identical pipeline to both CW-All and CW-HCG and compares the two ensemble populations directly.

\subsection{The COSMOS-Web weak-lensing mass calibration}
\label{sec:wl_calibration}

Halo masses in this catalog are anchored to a weak-lensing calibrated mass--luminosity relation,
$M$--$L_{\rm X}$, measured by \citet{Leroy2026} (in prep.) directly in the COSMOS-Web field from stacked
weak-lensing shear profiles of the same AMICO-selected group population characterized here (Eq.~\ref{eq:ML}
below). We summarize the calibration procedure here; full details are given in that paper.

The weak-lensing source sample is drawn from a shape catalogue measured independently in the
JWST/NIRCam F115W and F150W COSMOS-Web mosaics with \texttt{pyRRG}
\citep[following the moments-based PSF-corrected shear estimator of][]{Rhodes2000}, using
exposure-based WebbPSF models to characterize the spatially varying point-spread function. After
quality cuts (finite, positive shear-component errors; a valid photometric redshift; at least two
contributing exposures; $20.0 \leq {\tt MAG\_AUTO} \leq 27.5$; $|\gamma_1|,|\gamma_2|<1$), the
fiducial source catalogue contains 165,577 galaxies. For each lens, background sources are selected
with a conservative photometric-redshift cut ($z_{\rm phot} > z_{\rm l}+0.1$, and $z_{\rm l68} >
z_{\rm l}+0.1$ where the lower 68\% redshift bound is available) to minimize dilution of the
tangential shear by foreground and lens-associated galaxies.

Unlike the fiducial $M$--$L_{\rm X}$ relation itself, which is applied to every individually
detected group regardless of background-quality tier (Sect.~\ref{sec:catalog_overview}), the lens
sample used to \emph{calibrate} that relation is restricted to a cleaner subset: individually
detected CW-All groups with GOOD or NORMAL background quality that are neither flagged as
projected-contaminated nor as suspected false positives (365 of 499 CW-All detections, 73\%) lying
within the weak-lensing footprint, away from its edges and masked regions. These groups are partitioned into an iteratively repaired grid of luminosity--redshift
stacking bins (11 stacked bins of 8--23 systems each, spanning $0.05 \lesssim z \lesssim 1.5$ and
$L_{\rm X}/E(z) \sim 4\times10^{41}$--$1.4\times10^{43}\,\mathrm{erg\,s^{-1}}$, plus two individually
luminous, high-mass singleton systems retained as diagnostic validation points at $z=0.71$ and
$z=1.27$), designed to keep each stack populated while avoiding bins that mix physically distinct
regions of the $L_{\rm X}/E(z)$--$z$ plane. For each bin, the stacked tangential excess surface
density $\Delta\Sigma(R)$ is fit with an NFW halo model \citep{Navarro1997} over $0.3$--$2$~Mpc,
marginalizing over a Rayleigh-like miscentring-offset prior (characteristic scale
$\sim$0.1--0.3~Mpc) to account for possible offsets between the X-ray-derived group center and the
true lensing-mass centroid, with concentration fixed via a Duffy-like mass--concentration relation
\citep{Duffy2008} rather than fit jointly with mass. The resulting stacked $M_{200}$ posteriors are
then regressed against $L_{\rm X}/E(z)$ in log space with \texttt{linmix} \citep{Kelly2007}, which
accounts for uncertainties in both the lensing mass and the luminosity proxy, yielding the
$A_{\rm ML}$, $\alpha_{\rm ML}$, and $\sigma_{\rm int}$ values adopted in Eq.~\ref{eq:ML} below. This
is a single relation fit jointly over the full $0.05 \lesssim z \lesssim 1.5$ range of the lens
sample (via the $E(z)$ scaling already built into both axes), not separate relations fit in bins of
redshift; the sample size does not yet support splitting the weak-lensing calibration itself into
independent redshift bins. We
adopt this relation as fiducial over the earlier COSMOS calibration of \citet{Leauthaud2010}, using
it directly on the group population and X-ray pipeline presented in this paper and with higher
weak-lensing depth and resolution than was available to that earlier work (JWST versus ground-based
imaging); we compare the two calibrations quantitatively in Sect.~\ref{sec:scaling}.

\subsection{Halo mass from X-ray luminosity and temperature}
\label{sec:scaling}

We release both a luminosity-based and a temperature-based halo mass for every group, not as two
independent cross-checks, but because the temperature-based estimate is itself derived through
$L_{\rm X}$ (Eq.~\ref{eq:TL} below) and is constructed to agree with the luminosity-based estimate
by construction (Eq.~\ref{eq:MT}): the two are kept self-consistent so that users needing a
temperature alongside a mass (e.g.\ for gas physics or entropy arguments) are not forced to combine
mismatched scaling relations, while users who need only mass can use either column
interchangeably. The primary relation is
the weak-lensing calibrated mass--luminosity relation of \citet{Leroy2026} (in prep.) described in
Sect.~\ref{sec:wl_calibration}, which supersedes the earlier \citet{Leauthaud2010} relation used in
previous COSMOS group studies:
\begin{equation}
\log_{10}\left[\frac{M_{200}\, E(z)}{10^{14}~M_\odot}\right] = A_{\rm ML} + \alpha_{\rm ML}
\log_{10}\left[\frac{L_{\rm X}\, E(z)^{-1}}{10^{43}~{\rm erg~s}^{-1}}\right],
\label{eq:ML}
\end{equation}
with $A_{\rm ML} = 0.138^{+0.143}_{-0.160}$, $\alpha_{\rm ML} = 0.759^{+0.257}_{-0.220}$, and an
intrinsic scatter of $\sigma_{\rm int} = 0.133^{+0.104}_{-0.066}$~dex; equivalently, $M_{200}\,E(z) =
1.37^{+0.53}_{-0.42} \times 10^{14}~M_\odot\,[L_{\rm X}/E(z)/10^{43}~{\rm erg~s}^{-1}]^{0.759}$.
Reported $M_{200}$ uncertainties combine, in quadrature, the propagated $L_{\rm X}$ measurement
error with the calibration uncertainty on $A_{\rm ML}$ and $\alpha_{\rm ML}$ above (symmetrized from
their quoted asymmetric bounds); the intrinsic scatter $\sigma_{\rm int}$ is a separate,
population-level quantity and is not folded into the per-group uncertainty.

\citet{Leauthaud2010} calibrated the previous COSMOS $M$--$L_{\rm X}$ relation from Subaru weak
lensing combined with the earlier XMM-COSMOS group catalog of \citet{Finoguenov2007}, finding a
group-only slope of $\alpha \approx 0.66 \pm 0.14$ and a groups+clusters slope of $\alpha \approx
0.64 \pm 0.03$ in the same $M$--$L_{\rm X}$ direction used here, over a mass range limited to
$M_{200} \gtrsim 2\times10^{13}~M_\odot$ ($\log_{10} M_{200} \gtrsim 13.3$). Our slope,
$\alpha_{\rm ML} = 0.759^{+0.257}_{-0.220}$, is nominally steeper but statistically consistent with
both Leauthaud et al. values within the combined uncertainties, and, being measured directly with
deeper JWST weak lensing on the group population and X-ray pipeline used in this paper, extends the
direct calibration to more than an order of magnitude lower mass than the earlier work
(Sect.~\ref{sec:wl_calibration}).

For the majority of sources we do not have direct temperature measurements; we therefore estimate
the intragroup medium temperature from the rest-frame soft-band luminosity using the bias-corrected
$L_{\rm X}$--$T_{\rm X}$ relation from \citet{Kettula2015} (CFHTLenS weak lensing calibrated), where
$E(z) = \sqrt{\Omega_{\rm m}(1+z)^3 + \Omega_\Lambda}$ is the dimensionless Hubble parameter for our
adopted cosmology,
\begin{equation}
T_{\rm X} = T_0 \left( \frac{L_{\rm X}\, E(z)^{-1}/L_0}{A_{\rm LT}} \right)^{1/\alpha_{\rm LT}},
\label{eq:TL}
\end{equation}
with $T_0 = 5$~keV, $L_0 = 10^{44}$~erg~s$^{-1}$, $A_{\rm LT} = 1.51$, and $\alpha_{\rm LT} = 2.52$.
All temperatures reported in this catalog are therefore derived from this relation, not from
independent spectral fits. We assign a fractional uncertainty of 50\% to the luminosity-based
temperature estimate to reflect the intrinsic scatter of the relation and the lack of multi-band
constraints. The \citet{Kettula2015} calibration sample is itself dominated by group- and
cluster-scale systems ($\log_{10} M_{200}/M_\odot \gtrsim 13.5$); simulation-based $M$--$T$ relations
(e.g.\ Simba-C) suggest the true relation flattens below this range, so temperatures extrapolated
for our lowest-mass, coolest systems are more uncertain than the quoted 50\% scatter alone would
suggest, and should be treated as order-of-magnitude estimates rather than precision measurements
for that subpopulation.

For the conversion from temperature to mass we use an $M$--$T$ relation derived from the same
\citet{Leroy2026} (in prep.) $M$--$L_{\rm X}$ relation and the \citet{Kettula2015} $L_{\rm X}$--$T$ relation
above, so that the chain $L_{\rm X} \to T \to M$ agrees with $L_{\rm X} \to M$ directly:
\begin{equation}
M_{500} = A_{500} \left(\frac{T}{T_0}\right)^{\alpha} E(z)^{-1} \times 10^{14}~M_{\odot},
\label{eq:MT}
\end{equation}
with $\alpha = \alpha_{\rm ML}\,\alpha_{\rm LT} \approx 1.91$ and $A_{500} \approx 7.70$. We convert
to $M_{200}$ via $M_{200} = 1.4\,M_{500}$ (NFW concentration $c \approx 4$; \citealt{Navarro1997}),
and characteristic radii follow from $R_{\Delta} = \left[3 M_{\Delta}/(4\pi \Delta
\rho_c(z))\right]^{1/3}$.

\section{X-ray Properties of the Group Samples}
\label{sec:catalog_overview}

This section presents descriptive statistics of the released catalogs; detailed physical
interpretation is deferred to companion and future papers (Appendix~\ref{app:release}).

\subsection{Redshift distribution}
\label{sec:redshift_dist}

\begin{figure}[htbp]
\centering
\includegraphics[width=\linewidth]{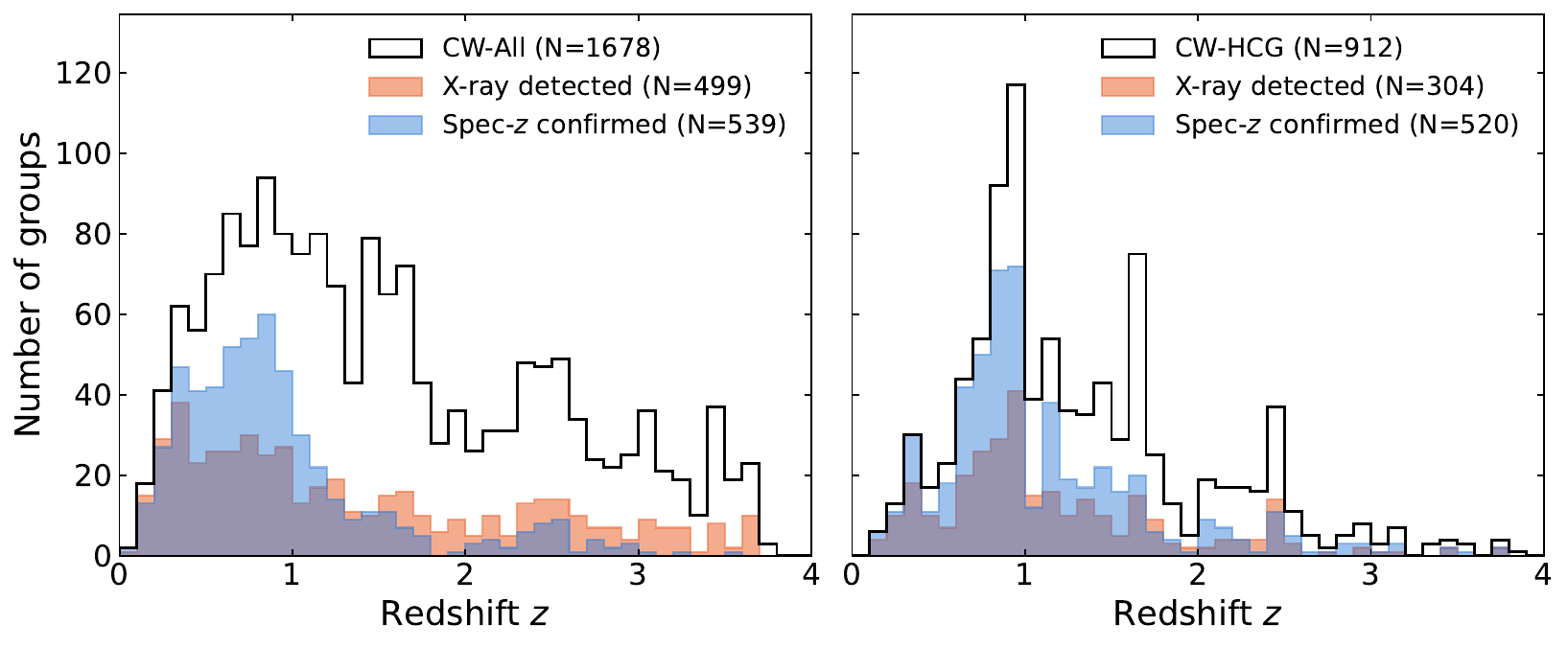}
\caption{Redshift distributions of the CW-All (left, N=1678) and CW-HCG (right, N=993) samples, each
showing the full catalog (black outline), the X-ray-detected subsample (orange; N=499 CW-All, N=332
CW-HCG), and the spectroscopically-confirmed subsample (blue; at least one member with a secure
spectroscopic redshift, $dz \leq 0.003$, from our probabilistic $P_z\times P_v$ membership pipeline,
Sect.~\ref{sec:membership}; N=539 CW-All, N=569 CW-HCG), on a
common $y$-axis scale for direct comparison.}
\label{fig:redshift_dist}
\end{figure}

Fig.~\ref{fig:redshift_dist} presents the redshift distributions of the CW-All and CW-HCG samples
side by side. CW-All spans $0.08 \lesssim z \lesssim 3.70$ with a median redshift of $z \approx
1.1$; CW-HCG spans $0.08 \lesssim z \lesssim 3.82$ with a median redshift of $z \approx 1.13$. Both
panels show the full catalog together with the X-ray-detected and spectroscopically-confirmed
subsamples; all three distributions peak at $z \sim 1$--1.5 in both catalogs, and the X-ray-detected
and spec-$z$ subsamples fall off more steeply toward high redshift than the parent sample, reflecting
surface-brightness dimming and spectroscopic depth, respectively. CW-HCG shows a more sharply peaked
distribution around $z \approx 0.9$--1.0 than the broader CW-All distribution, dropping by a factor
of $\sim$2 in the following bin ($1.0 \leq z < 1.1$). We attribute this to the spectroscopic
compilation itself rather than to the group-finding algorithms: the underlying spec-$z$ sampling
density in the COSMOS field peaks at $z \approx 0.9$--1.0 and falls by a comparable factor
immediately beyond it, a well-known consequence of strong rest-frame optical emission lines
([O\,\textsc{ii}], [O\,\textsc{iii}], H$\alpha$) shifting through the observed optical window at
this redshift. Because Hickson-style compact-group identification depends more heavily on secure
velocity confirmation than the photometric-redshift-driven AMICO algorithm, CW-HCG inherits this
spec-$z$ availability peak more strongly than CW-All.

\subsection{X-ray detection statistics}

\begin{table*}
\caption{Detection summary for COSMOS-Web group samples (SNR $\geq 2.0$). Median properties are computed for the
  detected subsample. The CW-All sample is the primary focus of this paper (Paper~I); a full analysis of CW-HCG is
  presented in Paper~II (Gozaliasl et al., in preparation). CW-HCG statistics here are included for reference only.}
\label{tab:detection}
\centering
\small
\setlength{\tabcolsep}{4pt}
\begin{tabular}{lrrrccccc}
\hline\hline
Catalog & $N_{\rm tot}$ & $N_{\rm det}$ & Det.\ (\%) & $\tilde{f}_{\rm X}$ &
$\widetilde{\log L_{\rm X}}$ & $\tilde{z}$ & $\tilde{M}_{200}^{L_{\rm X}}$ & $\tilde{R}_{200}^{L_{\rm X}}$ \\
 & & & & (erg cm$^{-2}$ s$^{-1}$) & [erg s$^{-1}$] & & ($M_{\odot}$) & (kpc) \\
\hline
CW-All  & 1678 & 499 & 29.7 & $6.82 \times 10^{-16}$ & 42.78 & 1.09 & $3.0 \times 10^{13}$ & 381 \\
CW-HCG$^{a}$ & 993  & 332 & 33.4 & $6.29 \times 10^{-16}$ & 42.69 & 0.95 & $2.7 \times 10^{13}$ & 411 \\
\hline
\end{tabular}
\tablefoot{$^{a}$ CW-HCG detection statistics are listed for comparison; the full CW-HCG X-ray analysis, stacking
  analysis, and comparison with CW-All are presented in Paper~II.}
\end{table*}

Table~\ref{tab:detection} summarizes the detection statistics for CW-All and CW-HCG. Throughout this
paper, "detection" refers to a significant ($\mathrm{SNR}\geq2.0$) source in the fixed-aperture
X-ray photometry measured at each group's optically determined position (Sect.~\ref{sec:analysis}),
not a blind, independent X-ray source search. We detect X-ray
emission from 499 of 1,678 CW-All groups (29.7\%) and 332 of 993 CW-HCG groups (33.4\%); detected
systems cluster around $z \sim 1$ with median luminosity $\log_{10}(L_{\rm X}/\mathrm{erg~s}^{-1})
\approx 42.7$--42.8 and median halo mass $M_{200} \approx 2.7$--$3.0 \times 10^{13}~M_{\odot}$ in
both catalogs. Ensemble properties of the non-detected majority, recovered via X-ray stacking, are
presented in Paper~II. This detected sample is substantially larger than earlier COSMOS X-ray group
compilations at comparable depth: \citet{Gozaliasl2019} identified 247 X-ray groups over $0.08 \leq
z < 1.53$ at the same effective flux limit ($F_{\rm lim} \approx 3\times10^{-16}~\mathrm{erg~cm^{-2}~s^{-1}}$), an order of magnitude deeper than \citet{George2011}; the 803 individually
detected CW-All and CW-HCG groups presented here, extending to $z \approx 3.8$, more than triple that
detected sample at the same depth, reflecting the larger input group catalogs and wider redshift
reach enabled by JWST-based selection.

To illustrate the brightest end of both X-ray samples directly in this catalog paper,
Table~\ref{tab:xray_sample_all} and Table~\ref{tab:xray_sample_hcg} (Appendix~\ref{app:release})
list the 20 highest-significance individually detected CW-All and CW-HCG systems, respectively.
These objects span the full redshift range of the compact-group sample and provide a concise view
of the strongest detections that anchor the
released catalog.

Fig.~\ref{fig:upper_limits} presents the upper-limit quality diagnostics for both catalogs.
The left panels show the flux distribution of all non-detected groups as 95\% Gaussian upper
limits (gray), overlaid on the flux distribution of detections (blue). A subset of upper
limits (orange) fall below the first quartile of the detected flux distribution; these are flagged
as likely measurement artifacts (e.g.\ from heavily masked apertures or low-coverage pointings)
rather than physically faint intragroup emission, and are excluded from stacking analyses.
The right panels show the redshift distribution of flagged versus unflagged non-detections:
flagged groups occur preferentially at low redshift where the X-ray mosaic has the deepest
coverage and aperture masking is most aggressive, consistent with a measurement rather than
astrophysical origin. Removing these flagged upper limits changes the stacked fluxes by less
than the statistical uncertainties and does not affect any conclusion in this paper.

\begin{figure*}[htbp]
\centering
\includegraphics[width=0.3\textwidth]{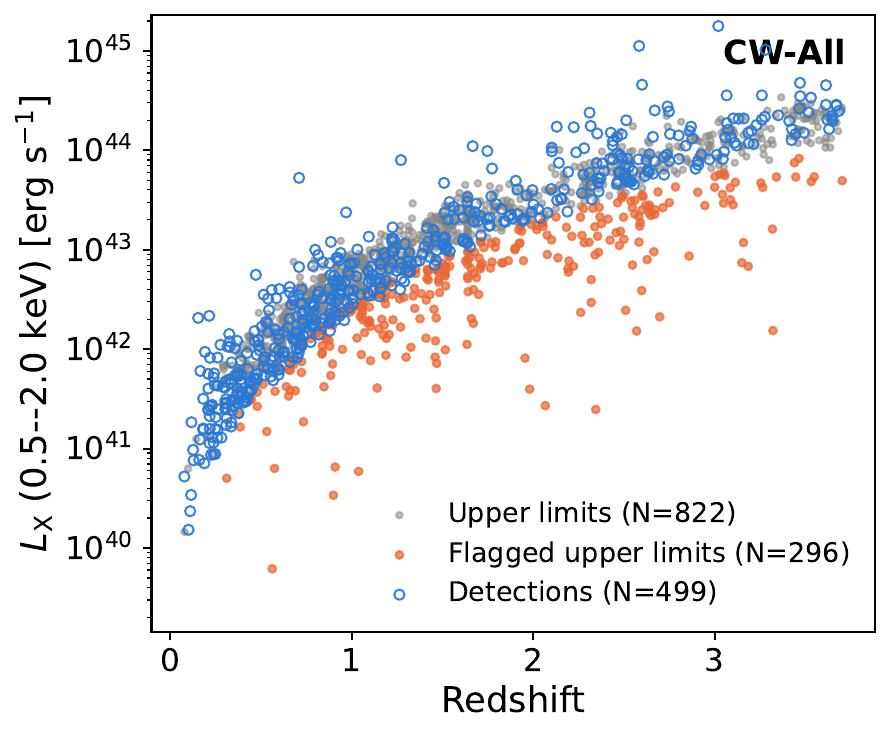}
\includegraphics[width=0.3\textwidth]{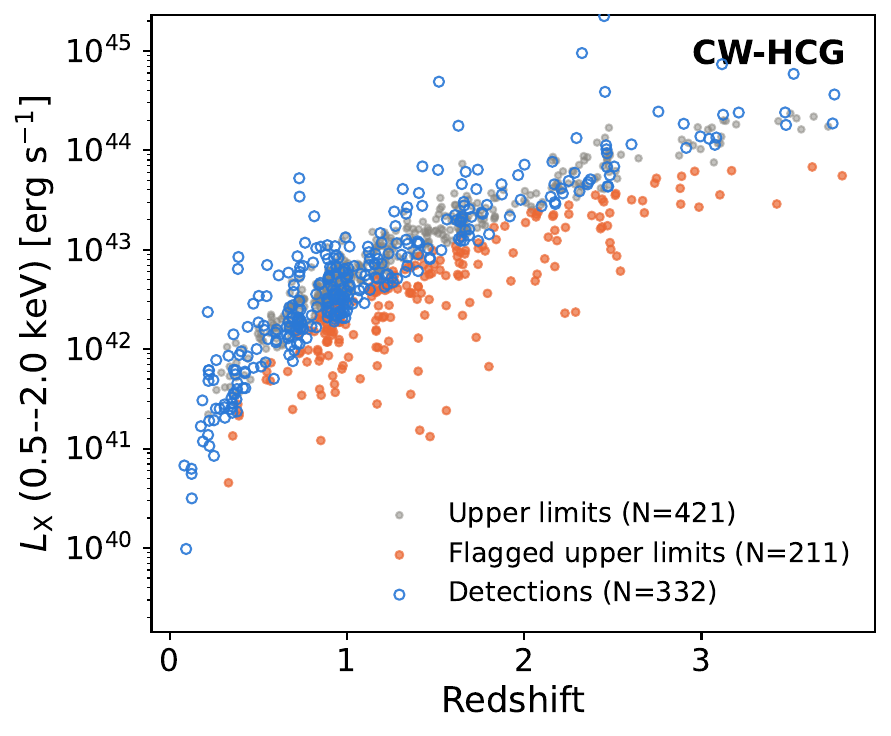}
\includegraphics[width=0.39\textwidth]{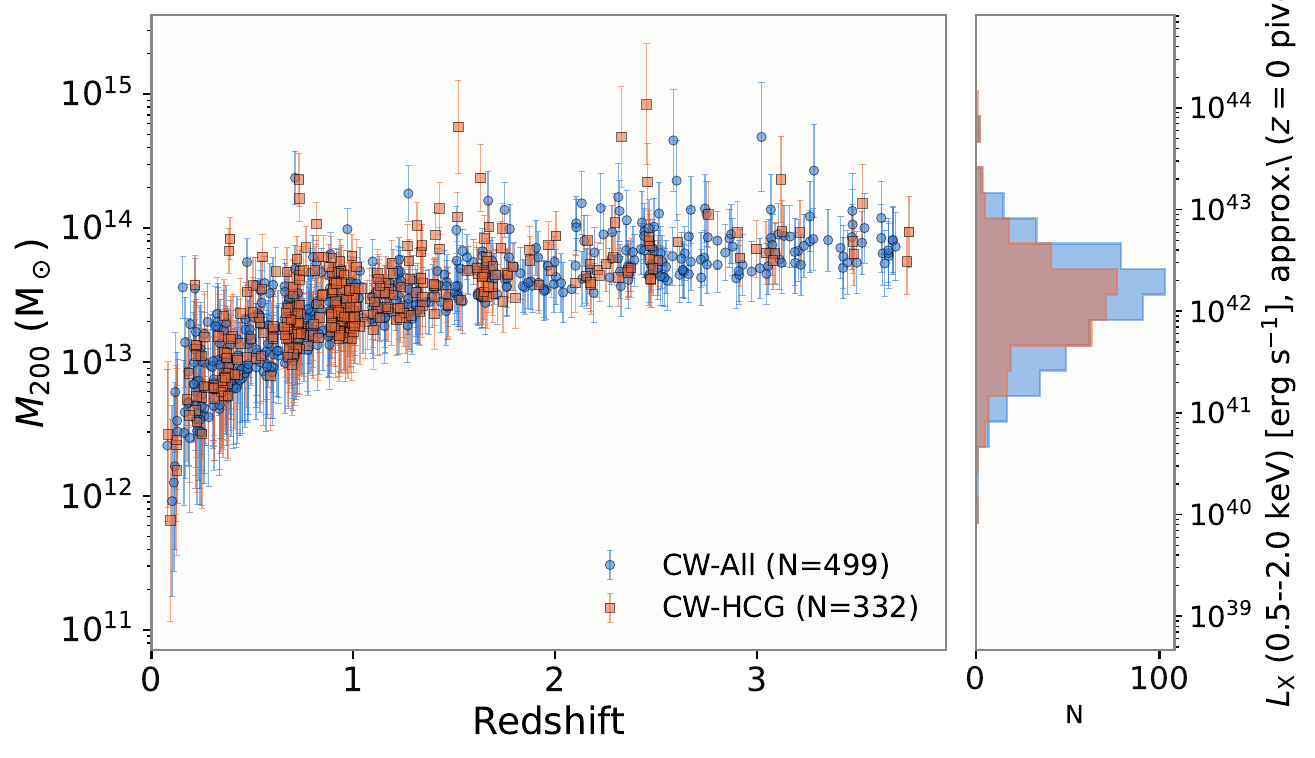}
\caption{Left, middle: rest-frame 0.5--2.0~keV luminosity $L_{\rm X}$ as a function of redshift for
CW-All (left) and CW-HCG (middle). Blue open circles: individual detections (SNR~$\geq 2.0$). Grey
points: non-detections with 95\% Gaussian upper limits; orange points: upper limits below the first
quartile of the detected flux distribution (flagged). This is the primary diagnostic for X-ray
luminosity across both samples (see Sect.~\ref{sec:catalog_overview}). Users requiring a
higher-confidence subsample can additionally cut on the released SNR column (e.g.\ SNR~$\geq 2.5$).
Right: the corresponding halo mass $M_{200}^{L_{\rm X}}$ versus redshift for detected groups in
CW-All (blue circles) and CW-HCG (orange squares), using the weak-lensing calibrated $M$--$L_{\rm
X}$ scaling relation of \citet{Leroy2026} (in prep.); note that $M_{200}^{L_{\rm X}}$ is derived directly
from $L_{\rm X}$ via a fixed power law and is therefore not an independent mass proxy (see
Sect.~\ref{sec:selection_effects}). The secondary (right-hand) axis on the marginal $M_{200}$
histogram gives the corresponding $L_{\rm X}$ scale, via the same relation evaluated at a fixed
$z=0$ ($E(z)=1$) pivot for reference; because the true relation carries an additional $E(z)$
factor, this axis is only an approximate guide and not an exact per-object conversion.}
\label{fig:LX_z}
\end{figure*}

Fig.~\ref{fig:LX_z} shows the resulting $L_{\rm X}$--redshift plane for both CW-All and
CW-HCG: individual detections populate the upper envelope of the distribution, while
non-detections appear as 95\% upper limits (with the flagged subset highlighted in orange).
The rightmost panel of the same figure shows the corresponding $M_{200}$--redshift plane for
the detected subsample of both catalogs, illustrating that both selections predominantly probe
group-scale halos ($M_{200} \sim 10^{13}$--$10^{14}\,M_\odot$) across the full redshift range
of the survey.

At fixed redshift, the detected and (unflagged) upper-limit populations in Fig.~\ref{fig:LX_z}
overlap substantially in $L_{\rm X}$ rather than being cleanly separated at a single threshold
luminosity. At fixed aperture size and
within a single narrow redshift slice, the net-count uncertainty $\sigma_{\rm net}$ among
non-detected groups spans a factor of $\sim$2 between its 5th and 95th percentiles survey-wide, and
the background level itself spans a comparably wide, occasionally negative (noise-dominated) range;
since $\mathrm{SNR}=C_{\rm net}/\sigma_{\rm net}$ (Sect.~\ref{sec:aperture}), two groups of the same
true $L_{\rm X}$ can therefore fall on either side of our SNR~$\geq2.0$ threshold depending on the
local background level and effective exposure where they happen to lie on the combined
Chandra+XMM-Newton mosaic (vignetting, CCD-gap and chip-edge coverage), not solely on their
intrinsic luminosity. A second contribution is the two-pass aperture scheme
of Sect.~\ref{sec:aperture} itself: non-detections retain the fixed first-pass 300~kpc aperture,
while detections are re-measured in the second pass with an $R_{500}$/$R_{200}$-scaled aperture that
is systematically larger for more luminous, higher-mass systems, so the two populations are not
measured with strictly comparable apertures even before any spatial noise variation is considered.
The $1/V_{\rm max}$ XLF of Sect.~\ref{sec:xlf_estimator} is built from
detections only, using a single, survey-wide flux limit $F_{\rm lim}$ rather than a
position-dependent completeness function; as already noted there, this is an intentional
simplification for the descriptive, catalog-level XLF presented here, but the overlap visible in
Fig.~\ref{fig:LX_z} is precisely the symptom of that simplification, and it means the true XLF
shape could differ from the one reconstructed here, most plausibly at the faint end of each
redshift bin where the detected/non-detected boundary is least sharp: sources scattered below
threshold by locally elevated background are simply absent from the $1/V_{\rm max}$ sum, while nothing
in the estimator adds their upper-limit information back in. We do not attempt a joint
detection-plus-censored-upper-limit likelihood (e.g.\ Kaplan-Meier or a full survival-analysis
treatment) in this catalog paper; doing so properly would require exactly the position- and
redshift-dependent completeness function $C(F,z)$ already flagged in Sect.~\ref{sec:xlf_estimator}
as necessary before the XLF shape is used for precision inference, and we defer it to that future,
forward-modeled treatment rather than introduce an ad hoc correction here.

Appendix~\ref{app:upper_limits_fig} (Fig.~\ref{fig:upper_limits}) shows the full upper-limit
diagnostic plots underlying this summary.

\subsection{High-redshift X-ray groups and a compact-group showcase: CW-HCG~01101}
\label{sec:showcase}

Both released catalogs extend the individually detected X-ray group population well beyond $z=2$.
Of the 505 CW-All systems at $z\geq2.0$, 133 (26.3\%) are individually X-ray detected, and of the
177 CW-HCG systems at $z\geq2.0$, 43 (24.3\%) are detected; the single highest-redshift individual
detection in either catalog reaches $z\approx3.7$--3.8. Restricting to clean detections (excluding
groups flagged as projection-contaminated or as suspected false positives) leaves 58 CW-All and 18
CW-HCG systems at $z\geq2.0$, with luminosities spanning $\log_{10}(L_{\rm X}/\mathrm{erg~s}^{-1})
\approx 43.8$--44.7 and halo masses $\log_{10}(M_{200}/M_\odot)\approx13.5$--14.4. These high-$z$
detections are necessarily biased toward the most luminous, highest-mass systems at fixed redshift
(surface-brightness dimming and cosmological dimming of the flux; Sect.~\ref{sec:selection_effects}),
but they demonstrate that group- and cluster-scale X-ray halos are already assembled and observable
with the current Chandra+XMM-Newton depth well into the first $\sim2.5$~Gyr of cosmic history.

As an illustration of an individual detection in this high-$z$ regime, Appendix~\ref{app:showcase}
presents CW-HCG group 01101, a Hickson-style compact group at $z=2.508$ individually detected at
SNR~$=3.4$, in the same region of COSMOS as the distant X-ray cluster reported by
\citet{Wang2016}.

As a further test of the high-$z$ population, we searched for diffuse, structure-scale X-ray
emission associated with ``AmicOne,'' the largest protocluster candidate identified by
\citet{Toni2025} from a three-dimensional clustering analysis of AMICO detections spanning $2.5 \leq
z \leq 3.09$. Using both a per-group/tile stacking approach and a membership-free cross-correlation
against the projected galaxy overdensity, we find no statistically significant diffuse signal at the
current depth (whole-box $3\sigma$ upper limit $\log_{10}(L_{\rm X}/\mathrm{erg~s}^{-1}) < 44.12$ at
$z=2.65$); we regard this as a depth-limited non-detection rather than evidence against AmicOne
itself, since surface-brightness dimming alone could place genuine, still-forming intragroup gas
below our sensitivity even for a real structure. The full search, null tests, and caveats are
presented in Appendix~\ref{app:amicone}.

\section{X-ray Luminosity Function}
\label{sec:xlf}

This is the primary statistical tool for characterizing a group and cluster population as a whole:
it links directly to the halo mass function through the $L_{\rm X}$--$M$ scaling relation
(Sect.~\ref{sec:scaling}) and provides the benchmark against which evolutionary models of
intragroup/intracluster gas heating and enrichment are tested \citep{Mullis2004,Vikhlinin2009}. We
present and discuss the XLF itself in this section; converting it into a group-scale halo mass
function via the same $M$--$L_{\rm X}$ relation -- which would ease direct comparison with the
literature halo mass function and let the non-linear shape of that relation be used to separate the
relative amount of evolution in $L^\star$ from that in the characteristic halo mass $M_h^\star$ -- is
deferred to a companion paper.
Uniform group catalogs with consistent X-ray measurements across a wide redshift baseline, such as
those released here, are needed to trace this evolution self-consistently within a single survey
rather than combining heterogeneous compilations. The current benchmark for wide-area group and
cluster demographics is the SRG/eROSITA All-Sky Survey \citep{Bulbul2024}, while the eROSITA Final
Equatorial-Depth Survey (eFEDS) provides the deepest eROSITA group-scale reference to date, reaching
$L_{\rm X} \approx 9\times10^{40}$--$4\times10^{44}~\mathrm{erg~s^{-1}}$ over $0.02 \lesssim z
\lesssim 0.94$ (542 systems, median $z=0.35$; \citealt{Liu2022,Bahar2022}). The X-ray pipeline
presented here reaches comparably faint fluxes ($F_{\rm lim} \approx 3\times10^{-16}~\mathrm{erg~cm^{-2}~s^{-1}}$) but over a much wider redshift baseline ($0.05 \lesssim z \lesssim 3.8$),
extending group-scale X-ray characterization beyond the eROSITA-accessible regime and into the
epoch where group-scale halos are still assembling. Sect.~\ref{sec:xlf_estimator} presents the
luminosity function of the individually X-ray-detected CW-All and CW-HCG groups;
Sect.~\ref{sec:xlf_schechter} gives the resulting Schechter fits and the evolution of the
characteristic luminosity $L^\star$; Sect.~\ref{sec:xlf_robustness} tests that evolution against
methodological choices; and Sect.~\ref{sec:selection_effects} assesses how much of the inferred
$L^\star$ evolution is driven by the flux-limited selection itself.

\subsection{$1/V_{\rm max}$ estimator}
\label{sec:xlf_estimator}

\begin{figure*}[t]
\centering
\includegraphics[width=\textwidth]{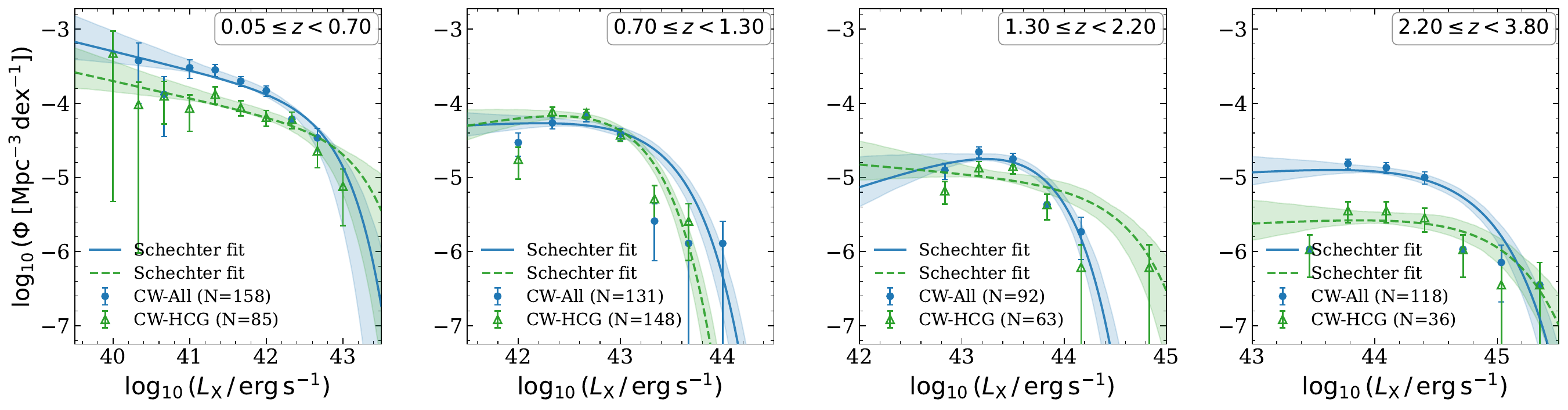}
\caption{X-ray luminosity function of COSMOS-Web groups in four redshift bins, derived
using the $1/V_{\rm max}$ estimator. Blue circles show CW-All; green circles show CW-HCG. Error bars
are the quadrature-summed $1/V_{\rm max}$ uncertainties per luminosity bin. Solid/dashed curves and
shaded bands show the Schechter fits and 68\% credible bands of Table~\ref{tab:xlf_schechter}. The
sampled luminosity range shifts progressively brighter from left to right as the flux limit excludes
fainter groups at higher redshift.}
\label{fig:xlf}
\end{figure*}

Fig.~\ref{fig:xlf} shows the rest-frame 0.5--2~keV X-ray luminosity function (XLF) of X-ray
detected COSMOS-Web groups in four redshift bins ($0.05 \leq z < 0.70$, $0.70 \leq z < 1.30$, $1.30
\leq z < 2.20$, and $2.20 \leq z < 3.80$), computed using the $1/V_{\rm max}$ estimator
\citep{Schmidt1968}. The bin edges were chosen, guided by the $M_{200}$--redshift distribution of
the detected sample (Fig.~\ref{fig:LX_z}, right panel), to keep a workable number of detections
per bin in both catalogs simultaneously; the original two-bin split ($z=1.3$, $z=2.2$) is retained
for the high-redshift half, while the low-redshift bin is further split at $z=0.70$, which balances
the CW-HCG counts (158/131/92/118 for CW-All and 85/148/63/36 for CW-HCG across the four bins,
respectively) without pushing any bin below the $\sim$30-object floor needed for a workable
three-parameter Schechter fit (Sect.~\ref{sec:xlf_schechter}). For each detected group $i$ with
luminosity $L_{{\rm X},i}$ at redshift $z_i$, we compute the maximum redshift $z_{{\rm max},i}$
within the bin at which the source would still be detectable above the effective flux limit $F_{\rm
lim} \approx 3 \times 10^{-16}$~erg~cm$^{-2}$~s$^{-1}$ (the 5th percentile of detected fluxes). The
accessible comoving volume is
\begin{equation}
V_{{\rm max},i} = \frac{\Omega}{4\pi}\left[V_{\rm c}(z_{{\rm max},i}) - V_{\rm c}(z_{\rm min})\right],
\end{equation}
with $\Omega = 0.45~\mathrm{deg}^2$ the effective survey area, and the XLF per luminosity bin of
width $\Delta\log L$ is
\begin{equation}
\Phi(L_{\rm X}) = \frac{1}{\Delta\log L}\sum_{i \in \Delta\log L} \frac{1}{V_{{\rm max},i}},
\label{eq:xlf}
\end{equation}
with uncertainties given by the quadrature sum of individual $1/V_{\rm max}$ contributions per bin.
This estimator assumes 100\% completeness above $F_{\rm lim}$ and treats the sensitivity limit as a
single survey-wide scalar; it does not model spatial variations in exposure, off-axis PSF
degradation, or source confusion, all of which would in principle require a position- and
redshift-dependent completeness function $C(F,z)$ folded into Eq.~\ref{eq:xlf} rather than a sharp cut. We
consider this simplification adequate for the descriptive, catalog-level XLF presented here
(Sect.~\ref{sec:xlf_schechter} tests the resulting $L^\star$ evolution against this and other
assumptions), but a forward-modeled selection function, ideally validated against
injection-recovery simulations, would be needed before the XLF shape itself is used for precision
cosmological or astrophysical inference in future work.

At $z < 0.70$ the XLF is measured over roughly three orders of magnitude in luminosity, from
$L_{\rm X} \sim 10^{40}$ to $10^{43}$~erg~s$^{-1}$, over which $\Phi$ declines steeply, consistent
with the power-law shape characteristic of the group-scale XLF \citep{Mullis2004,Vikhlinin2009}. At the faint end of this range ($L_{\rm X} \lesssim
10^{40.5}$--$10^{41}~{\rm erg~s}^{-1}$), the measured aperture flux is not guaranteed to be pure hot
intragroup-medium emission: unresolved X-ray binaries in the group's member galaxies contribute a
diffuse-looking floor that our point-source masking, built from the individually detected
Chandra-COSMOS-Legacy source catalog (Sect.~\ref{sec:analysis}), cannot remove, since an unresolved
XRB population is by construction below any individual point-source detection threshold. Using the
\citet{Lehmer2010} galaxy-wide $L_{\rm X}$--$M_\star$--SFR scaling relation ($L_{\rm X}(2$--$10~{\rm
keV}) = \alpha M_\star + \beta\,{\rm SFR}$, with $\alpha = 9.05\times10^{28}~{\rm
erg\,s^{-1}}\,M_\odot^{-1}$ and $\beta = 1.62\times10^{39}~{\rm erg\,s^{-1}\,(M_\odot\,yr^{-1})^{-1}}$)
and converting to our soft (0.5--2~keV) band with a representative XRB power-law spectrum
($\Gamma\approx1.7$), the aggregate unresolved-XRB contribution from a group's member population
with total stellar mass $10^{10}$--$3\times10^{11}~M_\odot$ and modest star formation is of order
$L_{\rm X} \sim 5\times10^{38}$--$2\times10^{40}~{\rm erg~s}^{-1}$ -- directly overlapping our
faintest reported detections. We do not attempt to model or subtract this contribution here, so
individual luminosities and upper limits below $L_{\rm X} \sim 10^{41}~{\rm erg~s}^{-1}$ should be
treated as upper bounds on the diffuse intragroup-medium emission specifically, rather than as pure
hot-gas measurements. This affects a small minority of individual detections (17 of 831 combined
CW-All plus CW-HCG detections, $\sim$2\%, lie below $L_{\rm X} = 10^{41}~{\rm erg~s}^{-1}$) and is
therefore unlikely to drive the Schechter fits of Sect.~\ref{sec:xlf_schechter}, but is directly
relevant to any use of the faintest individual entries in the public catalog
(Appendix~\ref{app:release}). CW-All and CW-HCG track each other closely at low luminosity
in every bin but separate increasingly toward the bright end and toward higher redshift, where the
compact-group subsample shows a lower space density, likely reflecting the smaller effective volume
sampled by the HCG selection combined with the shrinking number of detections per bin. Progressing
through the four bins, the range of luminosities sampled shifts steadily upward -- from $L_{\rm X}
\sim 10^{40}$--$10^{43}$ at $z<0.70$ to $L_{\rm X} \sim 10^{43}$--$10^{45}$~erg~s$^{-1}$ at $2.20
\leq z < 3.80$ -- as fainter groups fall below the flux limit; the overall amplitude decreases by
roughly an order of magnitude in $\Phi$ between the lowest and highest redshift bins. Throughout
this paper we describe evolution as a function of increasing redshift (i.e.\ back in cosmic time),
not of increasing cosmic time itself, so this decline in $\Phi$ toward higher redshift is what we
term \emph{negative density (amplitude) evolution}, to distinguish it explicitly from
the \emph{positive luminosity evolution} in $L^\star$ discussed in Sect.~\ref{sec:xlf_schechter} --
in principle a luminosity function can shift toward higher characteristic luminosity, decline in
overall amplitude, or both simultaneously, and only the fitted Schechter parameters of
Table~\ref{tab:xlf_schechter} (rather than the raw, flux-limit-affected binned counts shown here)
cleanly separate the two effects.

This behavior is a direct consequence of the survey's flux-limited selection rather than evidence
for a physically shrinking group population: as redshift increases, an ever larger fraction of the
intrinsic group population falls below $F_{\rm lim}$, so the detected XLF necessarily probes only
the increasingly luminous, high-mass tail of the underlying distribution. This is precisely why we
complement the direct $1/V_{\rm max}$ estimate with the parametric Schechter fits below
(Sect.~\ref{sec:xlf_schechter}), which separate the amplitude and shape of the space density from
the changing luminosity range actually sampled at each redshift. In the two lowest-redshift bins,
where both catalogs are simultaneously well sampled, CW-All and CW-HCG agree at the level of the
statistical uncertainties, indicating that at this survey's depth compact-group selection does not
by itself suppress the space density of X-ray-luminous systems relative to the general group
population. In the two highest-redshift bins the two catalogs visibly diverge at the bright end (by
up to $\sim$1~dex in $\Phi$), which we attribute to small-number statistics and volume effects
(Sect.~\ref{sec:groupcat}) given the modest bin populations there (CW-HCG: 64 and 34 groups), but we
cannot rule out a genuine physical difference in group formation efficiency with the present
four-bin,
non-parametric estimator alone.

\subsection{Schechter fits and $L^*$ evolution}
\label{sec:xlf_schechter}

We fit each binned XLF with a Schechter function,
\begin{equation}
\Phi(L_{\rm X})\,d\log L_{\rm X} = \ln(10)\,\Phi^\star \left(\frac{L_{\rm X}}{L^\star}\right)^{\alpha+1}
  \exp\!\left(-\frac{L_{\rm X}}{L^\star}\right) d\log L_{\rm X},
\end{equation}
using an affine-invariant MCMC sampler (\texttt{emcee}; 128 walkers, 12{,}000 steps, 3{,}000
burn-in) with flat priors on $\log\Phi^\star$, $\log L^\star$, and $\alpha$.
Table~\ref{tab:xlf_schechter} lists the posterior medians and 68\% credible intervals for both
catalogs in all four redshift bins.

\begin{table}
\caption{Schechter function best-fit parameters (posterior median and 68\% credible interval) for the COSMOS-Web group
  XLF, from the $1/V_{\rm max}$-binned data of Fig.~\ref{fig:xlf} (primary four-bin scheme, above the first
  horizontal rule) and Fig.~\ref{fig:lstar_agn} (two coarser two-bin cross-checks, testing sensitivity to the
  specific bin edges rather than a physically motivated split: one at $z=1.30$, reusing a four-bin edge; and one
  at $z=2.00$). $N_{\rm det}$ is the number of individually detected groups contributing to each fit.}
\label{tab:xlf_schechter}
\centering
\setlength{\tabcolsep}{4pt}
\renewcommand{\arraystretch}{1.5}
\begin{tabular}{lccccc}
\hline\hline
Catalog & $z$ range & $\log\Phi^\star$ & $\log L^\star$ & $\alpha$ & $N_{\rm det}$ \\
 & & (Mpc$^{-3}$) & (erg~s$^{-1}$) & & \\
\hline
CW-All & $0.05$--$0.70$ & $-4.35^{+0.32}_{-0.27}$ & $42.74^{+0.25}_{-0.38}$ & $-1.25^{+0.15}_{-0.16}$ & 158 \\
CW-HCG & $0.05$--$0.70$ & $-4.80^{+0.37}_{-0.44}$ & $43.18^{+0.75}_{-0.57}$ & $-1.23^{+0.14}_{-0.14}$ & 85 \\
CW-All & $0.70$--$1.30$ & $-4.50^{+0.13}_{-0.13}$ & $43.29^{+0.11}_{-0.11}$ & $-0.92^{+0.16}_{-0.16}$ & 131 \\
CW-HCG & $0.70$--$1.30$ & $-4.27^{+0.11}_{-0.11}$ & $42.96^{+0.10}_{-0.10}$ & $-0.74^{+0.21}_{-0.21}$ & 148 \\
CW-All & $1.30$--$2.20$ & $-4.76^{+0.12}_{-0.11}$ & $43.57^{+0.12}_{-0.13}$ & $-0.54^{+0.29}_{-0.28}$ & 92 \\
CW-HCG & $1.30$--$2.20$ & $-5.48^{+0.31}_{-0.32}$ & $44.51^{+0.32}_{-0.29}$ & $-1.12^{+0.22}_{-0.22}$ & 63 \\
CW-All & $2.20$--$3.80$ & $-5.11^{+0.15}_{-0.15}$ & $44.64^{+0.11}_{-0.12}$ & $-0.90^{+0.20}_{-0.20}$ & 118 \\
CW-HCG & $2.20$--$3.80$ & $-5.81^{+0.26}_{-0.27}$ & $44.94^{+0.27}_{-0.24}$ & $-0.91^{+0.27}_{-0.27}$ & 36 \\
\hline
CW-All & $0.05$--$1.30$ & $-5.04^{+0.15}_{-0.14}$ & $43.53^{+0.14}_{-0.17}$ & $-1.48^{+0.05}_{-0.05}$ & 289 \\
CW-HCG & $0.05$--$1.30$ & $-4.52^{+0.09}_{-0.09}$ & $43.08^{+0.09}_{-0.09}$ & $-1.10^{+0.07}_{-0.07}$ & 233 \\
CW-All & $1.30$--$3.80$ & $-5.32^{+0.10}_{-0.10}$ & $44.64^{+0.09}_{-0.09}$ & $-1.03^{+0.08}_{-0.08}$ & 210 \\
CW-HCG & $1.30$--$3.80$ & $-6.25^{+0.24}_{-0.24}$ & $45.09^{+0.24}_{-0.22}$ & $-1.40^{+0.10}_{-0.11}$ & 99 \\
\hline
CW-All & $0.05$--$2.00$ & $-5.41^{+0.19}_{-0.17}$ & $43.89^{+0.18}_{-0.21}$ & $-1.55^{+0.05}_{-0.05}$ & 366 \\
CW-HCG & $0.05$--$2.00$ & $-5.25^{+0.13}_{-0.12}$ & $43.81^{+0.12}_{-0.13}$ & $-1.41^{+0.05}_{-0.05}$ & 289 \\
CW-All & $2.00$--$3.80$ & $-4.98^{+0.09}_{-0.09}$ & $44.50^{+0.08}_{-0.08}$ & $-0.57^{+0.16}_{-0.16}$ & 133 \\
CW-HCG & $2.00$--$3.80$ & $-6.00^{+0.29}_{-0.29}$ & $45.00^{+0.27}_{-0.24}$ & $-1.11^{+0.23}_{-0.22}$ & 43 \\
\hline
\end{tabular}
\renewcommand{\arraystretch}{1.0}
\end{table}

The fit range for each bin is restricted to the luminosity interval actually populated by detections
in that bin (Table~\ref{tab:xlf_schechter} and Fig.~\ref{fig:xlf} $x$-axes), rather than a single
shared range spanning the full survey: the two intermediate and high-redshift bins do not reach down
to $\log L_{\rm X} \sim 40$, so fitting them over an unnecessarily wide range would let the
Schechter shape be driven by empty bins rather than data. Across all four bins, the posterior-median
$\log L^\star$ rises monotonically with redshift in both catalogs -- from 42.74 to 44.64 in CW-All
($\sim$1.9 dex) and from 42.99 to 45.05 in CW-HCG ($\sim$2.1 dex) -- consistent with the negative
luminosity evolution seen directly in Fig.~\ref{fig:xlf}: at fixed number density, groups typical
of the highest-redshift bin are substantially more luminous than their low-redshift counterparts.
Whether this point-estimate trend constitutes statistically robust evolution, once the full (and in
places broad, asymmetric) MCMC posteriors are propagated rather than just their medians, is tested
quantitatively in Sect.~\ref{sec:xlf_robustness}. This is, as discussed above, the expected
signature of a flux-limited survey progressively losing faint groups rather than necessarily a
tracer of genuine evolution in the full underlying group population. The finer binning newly
resolves the trend as continuous and monotonic across all four steps rather than as a single jump
between two widely separated epochs, which was the most a two-bin scheme could show. The faint-end
slope $\alpha$ is, with the fit range restricted to the populated data, remarkably consistent
between the two catalogs at every redshift (Table~\ref{tab:xlf_schechter}), and itself evolves non-monotonically with redshift, flattening from $\alpha \approx
-1.2$ at low $z$ to $\alpha \approx -0.5$ at $1.30 \leq z < 2.20$ before steepening again at the
highest redshifts. Given the modest and uneven number of detected groups per bin (34--156) and the
correspondingly broad, sometimes asymmetric credible intervals -- most visible for CW-HCG at $0.05
\leq z < 0.70$ and $2.20 \leq z < 3.80$, its two thinnest bins -- we do not treat this non-monotonic
$\alpha(z)$ trend as a settled result, but the close agreement between the two
independently-selected catalogs at every redshift is a reassuring internal consistency check on the
fits themselves. We caution against over-interpreting any individual Schechter parameter beyond the
qualitative, monotonic $L^\star$ evolution and the good CW-All/CW-HCG agreement in $\alpha$.
Fig.~\ref{fig:lstar_agn} (Sect.~\ref{sec:selection_effects}) plots this faint-end slope
$\alpha(z)$ together with the $L^\star(z)$ evolution for both catalogs, in physical units.
Appendix~\ref{app:diagnostics} (Fig.~\ref{fig:xlf_corner}) shows
the full posterior for the CW-All, $0.70 \leq z < 1.30$ fit as a representative, well-constrained
example of the parameter correlations and credible regions underlying Table~\ref{tab:xlf_schechter};
corner plots for the remaining seven fits, including the unstable CW-HCG $2.20 \leq z < 3.80$ fit
discussed in Sect.~\ref{sec:xlf_robustness}, are available in the data release repository
accompanying this paper.

\subsection{Robustness of the $L^\star$ evolution}
\label{sec:xlf_robustness}

Because each Schechter fit has three free parameters and, in several bins, fewer than 100
detections, we test whether the monotonic $\log L^\star$ rise reported above is an artifact of
parameter degeneracy or of which particular groups happen to populate a bin, rather than a stable
feature of the data, using a fixed-$\alpha$ refit, a bootstrap over the detected group list, and the
full posterior-based significance of the rise between adjacent redshift bins
(Appendix~\ref{app:xlf_robustness}). The rise survives both alternative tests at every redshift step
except the lowest-redshift, CW-HCG-only step, which remains only marginal; the two steps in the
upper half of our redshift range are each significant beyond 95\% confidence
(Table~\ref{tab:xlf_significance}), while the full low-to-high-$z$ baseline comparison is consistent
with net positive evolution at close to the 95\% level in both catalogs. We conclude that the data
support genuine $L^\star$ evolution with high confidence in the upper half of our redshift range,
and are suggestive of, but do not tightly constrain, evolution in the lower half. As a further,
independent cross-check against the specific four-bin scheme, we repeat the same $1/V_{\rm max}$
and Schechter-fitting procedure using a single split at $z=1.30$ -- reusing, rather than
independently choosing, one of the four-bin edges above, so the two-bin scheme is simply the
$\{1,2\}$/$\{3,4\}$ merger of the same underlying bins (Table~\ref{tab:xlf_schechter},
Fig.~\ref{fig:lstar_agn}): both catalogs reproduce the same monotonic rise in $\log L^\star$ under
this much coarser binning, from $43.53\pm0.15$ to $44.64\pm0.09$ in CW-All and from $43.08\pm0.09$
to $45.09\pm0.23$ in CW-HCG, confirming that the evolution is not an artifact of the four-bin choice.
The two-bin CW-HCG faint-end slope is notably steeper than any of its four individual four-bin
values (Table~\ref{tab:xlf_schechter}); this is expected rather than a new tension, since its
high-redshift two-bin range ($1.30 \leq z < 3.80$) merges two sub-populations with very different
detected luminosity ranges (the $1.30 \leq z < 2.20$ and $2.20 \leq z < 3.80$ four-bin detections
span $\log L_{\rm X} \approx 42.8$--$43.9$ and $43.6$--$45.3$, respectively), so a single
non-evolving Schechter function fit to their union has its faint end pulled steep to accommodate a
luminosity break that is itself shifting within the bin -- the same effect that motivated the
finer four-bin scheme as our primary result.

As a second cross-check, independent of the specific four-bin edges, we repeat the same procedure
with a split at $z=2.00$ (Table~\ref{tab:xlf_schechter},
Fig.~\ref{fig:lstar_agn}). The monotonic rise persists under this split as well, from
$43.89\pm0.20$ to $44.50\pm0.08$ in CW-All and from $43.81\pm0.13$ to $45.00\pm0.26$ in CW-HCG.
Compared to the $z=1.30$ split, moving the boundary to $z=2.00$ pulls the CW-HCG high-redshift bin's
faint-end slope back toward its four-bin values ($\alpha = -1.11^{+0.23}_{-0.22}$, versus
$-1.40^{+0.10}_{-0.11}$ at $z=1.30$), because a smaller fraction of the very luminous, sparsely
sampled $z\geq2.2$ four-bin detections is now mixed into that bin -- consistent with the
luminosity-range-mixing explanation above rather than a new, independent tension.

We focus our evolution discussion on $L^\star$ and $\alpha$ rather than the Schechter normalization
$\Phi^\star$, and Fig.~\ref{fig:lstar_agn} shows $\Phi^\star(z)$ (panel c) primarily for
completeness rather than as a claim of comparable robustness. $\Phi^\star$ is degenerate with the
overall completeness/selection function at each redshift -- including the 100\%-above-$F_{\rm lim}$
simplification of Sect.~\ref{sec:xlf_estimator} and, more importantly for a CW-All-versus-CW-HCG
comparison, the differential spectroscopic-depth sensitivity between the two group-finders
(Sect.~\ref{sec:redshift_dist}, Fig.~\ref{fig:redshift_dist}): CW-HCG's identification depends more
heavily on secure velocity confirmation than AMICO's photometric-redshift-driven selection, so its
effective completeness (and hence its $\Phi^\star$) is more directly coupled to the spec-$z$
sampling density than CW-All's. $L^\star$ and $\alpha$, as shape parameters of the same fitted
function, are comparatively insulated from an overall normalization offset between the two catalogs
in a way that a direct $\Phi^\star$ comparison is not, which is why we treat $L^\star(z)$ and
$\alpha(z)$ as our primary, more robust results and report $\Phi^\star(z)$ as a secondary,
qualitative diagnostic. Bearing that caveat in mind, $\log\Phi^\star$ declines by
$\sim$0.5--1.5~dex across our four bins in both catalogs (Table~\ref{tab:xlf_schechter}), the
density-evolution counterpart to the positive $L^\star$ evolution discussed above (the
distinction between the two drawn explicitly in Sect.~\ref{sec:xlf_estimator}).

\subsection{Selection effects: is the $L^\star$ rise driven by mass selection?}
\label{sec:selection_effects}

A flux-limited survey unavoidably selects more luminous systems at higher redshift. Whether this
also means \emph{more massive} systems is not automatic: at fixed halo mass, higher-redshift systems
are more compact and can be intrinsically brighter in X-rays simply from higher gas density, AGN
feedback that has not yet had time to lower the core gas density can leave a halo of a given mass
X-ray brighter before it is regulated, and merger-driven boosts to the diffuse luminosity are more
common at earlier epochs -- so a rising $L^\star$ need not imply a rising typical host mass, and
could instead (or in addition) reflect these redshift-dependent astrophysical effects at fixed mass.
The halo mass $M_{200}^{L_{\rm X}}$ shown in Fig.~\ref{fig:LX_z} (right panel) is \emph{derived
from} $L_{\rm X}$ via the fixed \citet{Leroy2026} (in prep.) power law (Sect.~\ref{sec:scaling}) and therefore
cannot serve as an independent check of any of this -- by construction, $M_{200}^{L_{\rm X}}$
increases whenever $L_{\rm X}$ does. To test whether the $\sim$1.9--2.1~dex $\log L^\star$ rise
(Sect.~\ref{sec:xlf_schechter}) is simply a symptom of the detected population becoming more
massive with redshift, we instead examine mass/richness proxies that are independent of the
X-ray measurement. For CW-All we use both the AMICO group-finder's own intrinsic richness
$\lambda_\star$ \citep{Toni2025}, a validated group-scale mass proxy, and, as a cross-check, the
number of members assigned by our own probabilistic
$P_z\times P_v$ membership pipeline (Sect.~\ref{sec:membership}), which is entirely independent of
$\lambda_\star$. For CW-HCG we instead use $n_g$, Hasinger et al.'s own
discovery-time richness recorded in the parent compact-group catalog (Sect.~\ref{sec:groupcat}):
this is the group-finder's native richness statistic for CW-HCG, playing the same role that
$\lambda_\star$ plays for AMICO, and it is a cleaner and more directly interpretable choice than our
own membership-confirmed count, which for CW-HCG is dominated by the pipeline's redshift-consistency
gate rather than by the discovery selection itself. For both catalogs we also use the BGG stellar
mass (from SED fitting).

None of these proxies track the $L^\star$ evolution. The median AMICO richness $\lambda_\star$ of
X-ray-detected CW-All groups is 4.7, 8.0, 9.9, and 7.6 across the four bins -- essentially flat
within the bin-to-bin scatter, with no monotonic trend -- while the median membership richness of
the same X-ray-detected
CW-All groups is 13.0, 16.0, 11.0, and 5.5 members across the four bins -- a net \emph{decline} of
more than a factor of two from the second to the last bin. Both are the reverse of the
$\sim$1.9~dex rise in
$\log L^\star$ over the same range. CW-HCG's discovery richness $n_g$, by contrast, is essentially
flat at its selection floor: the median is exactly $n_g = 3$ in all four redshift bins, because 71\%
of the parent CW-HCG catalog has $n_g = 3$ (Sect.~\ref{sec:groupcat}) and this floor is preserved
among the X-ray-detected subsample at every epoch. $n_g$ is a static, discovery-time property of
each system and is not expected to track any redshift-dependent selection in the X-ray detection
itself, so its flatness here demonstrates that the X-ray-detected CW-HCG population is not
progressively drawn from intrinsically richer compact groups at higher redshift -- the detected
systems remain, by this measure, minimal-richness compact groups throughout. Taken together, the
CW-All decline and the CW-HCG flatness both point the same way: neither catalog shows the detected
population becoming richer with redshift, which is inconsistent with "richer/more massive systems at
high $z$" being the driver of the observed $L^\star$ evolution. The median BGG stellar mass is
comparably flat and non-monotonic in both catalogs ($\log M_\star^{\rm BGG} \approx 10.1$--$10.5$ in
all four CW-All bins and $10.15$--$10.43$ in CW-HCG, a total spread of $\lesssim$0.4~dex against a
$\sim$1.9~dex $L^\star$ rise). Fig.~\ref{fig:xlf_sample_comp} (Appendix~\ref{app:xlf_robustness})
shows these proxies together
with the number of detected groups per bin for both catalogs, so the CW-All $\lambda_\star$ and
membership-richness behavior, the
CW-HCG richness floor, and the flat BGG mass can be inspected directly alongside the sample sizes
entering each Schechter fit (Table~\ref{tab:xlf_schechter}). If the detected sample at high redshift
consisted mainly of systems hosted by richer, more massive halos and more massive central galaxies,
all three proxies should rise systematically with redshift; instead none does. We conclude that
mass/richness selection of the host population, as traced by these independent, X-ray-independent
proxies, is not the primary driver of the measured $L^\star$ evolution. We cannot rule out that the
X-ray-detected subset at fixed richness or BGG mass is itself a biased tracer of the underlying halo
population at each epoch (Eddington/Malmquist-type bias); disentangling this fully would require a
forward-modeled selection function (Sect.~\ref{sec:xlf}) beyond the scope of this catalog paper.

\begin{figure}[htbp]
\centering
\includegraphics[width=0.98\linewidth]{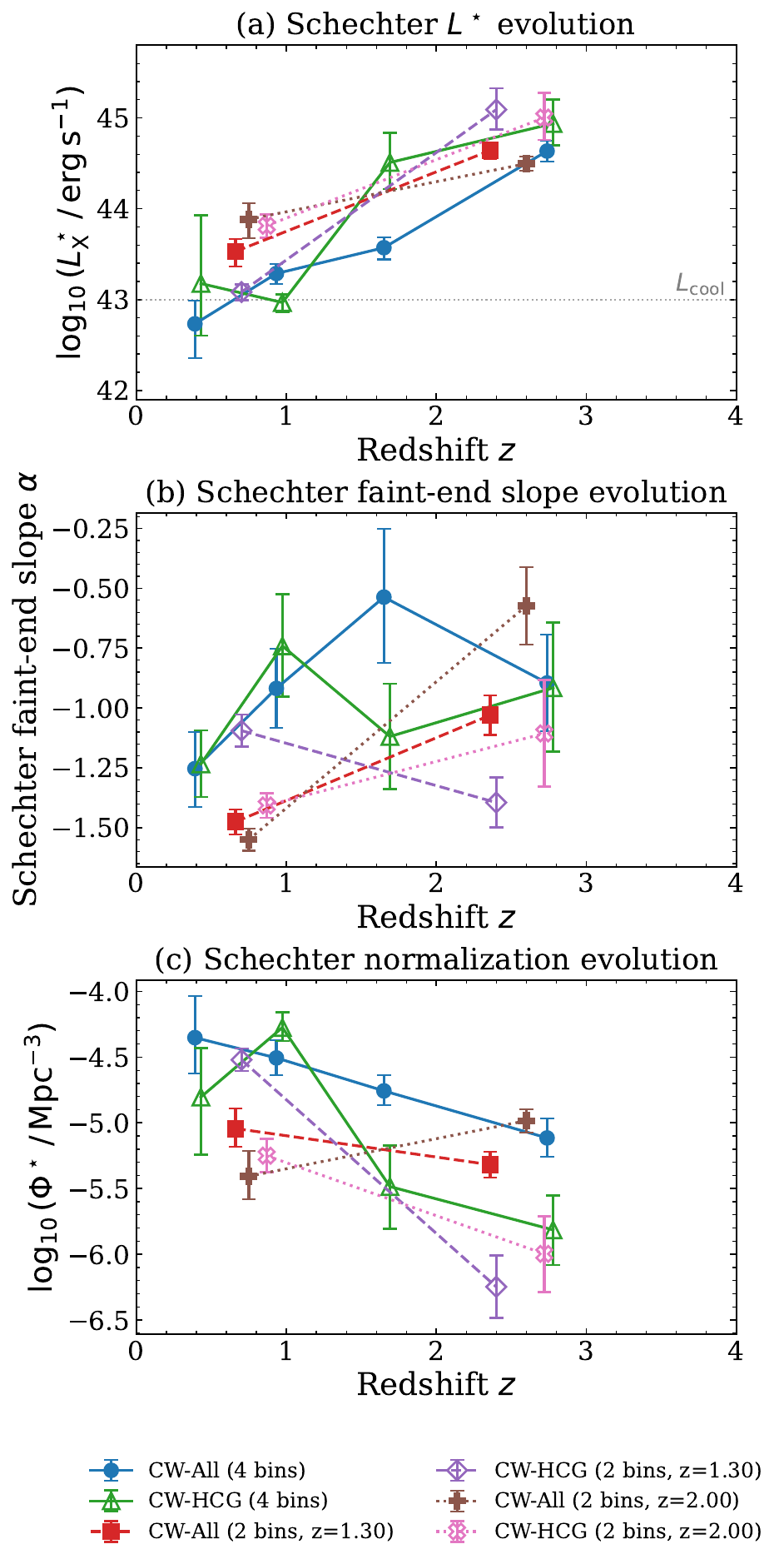}
\caption{Schechter evolution of the group XLF with redshift for CW-All (blue filled circles) and
CW-HCG
(green open triangles), from the four-bin fits of Table~\ref{tab:xlf_schechter}, together with two
coarser two-bin cross-checks, testing sensitivity to the specific bin edges rather than a physically
motivated split: one at $z=1.30$ (CW-All red filled squares, CW-HCG purple open diamonds), reusing a
four-bin edge, and one at $z=2.00$ (CW-All brown filled plus signs, CW-HCG pink open crosses).
(a) $\log L^\star(z)$, in
physical units (erg~s$^{-1}$); the grey dotted line marks $L_{\rm cool} \sim 10^{43}$~erg~s$^{-1}$,
an approximate threshold above which the intragroup medium cooling time becomes shorter than the
Hubble time. (b) Faint-end slope $\alpha(z)$. (c) Schechter normalization $\log\Phi^\star(z)$,
shown for completeness but interpreted with more caution than (a)/(b), since it is more directly
sensitive to the survey completeness/selection function than the shape parameters $L^\star$ and
$\alpha$ (Sect.~\ref{sec:xlf_schechter}). All three binning schemes reproduce the same
monotonic $L^\star$ rise in both catalogs.}
\label{fig:lstar_agn}
\end{figure}

Fig.~\ref{fig:lstar_agn} summarizes our own Schechter evolution: $\log L^\star(z)$ (panel a), the
faint-end slope $\alpha(z)$ (panel b), and the normalization $\log\Phi^\star(z)$ (panel c) for both
catalogs, in physical units. The X-ray emission underlying the group XLF (diffuse intragroup gas)
and that underlying the AGN/quasar luminosity function (nuclear accretion) are physically distinct
and a priori unrelated, but AGN feedback is the leading mechanism proposed to heat and quench
intragroup gas, so a coincidence (or lack thereof) between the cosmic AGN/quasar history and the
group XLF's own evolution is worth checking as a first-order consistency test, even though it cannot
demonstrate a causal link. We do not attempt a fit-to-fit comparison against the \citet{Hopkins2007}
AGN bolometric quasar luminosity function (QLF): the two luminosity classes differ (individual AGN
bolometric output versus diffuse group-scale soft X-ray gas emission, with a large and
physically-expected but scientifically uninformative absolute offset), and the two studies fit
different functional forms for the bright end (an exponential-cutoff Schechter function for our XLF
versus a double power law for the AGN QLF), so an $L^\star$ comparison would not be
guaranteed to mean numerically the same thing across the two fits. The one qualitative comparison
that avoids this is a check for a shared turnover: \citet{Hopkins2007}'s abstract reports that the
peak in the total bolometric quasar/AGN luminosity density is well constrained at $z = 2.15 \pm
0.05$, yet our own $\log L^\star_{\rm X}$ shows no corresponding turnover or feature, continuing to
rise monotonically through our highest-redshift bin ($2.20 \leq z < 3.80$, median $z \approx
2.5$--2.7 for CW-HCG and CW-All respectively), unambiguously past this epoch. We regard this as evidence against a simple reading in
which the group XLF passively tracks the cosmic AGN history, rather than as an inconclusive
non-detection of a link, but stress this is a single qualitative observation, not a demonstrated
physical connection: with only four $L^\star$ points, a correlation statistic would be dominated by
small-number statistics and by the fact that $\log L^\star_{\rm X}$ varies smoothly with redshift by
construction, so we do not compute one. A physical connection between AGN feedback and the
group-scale XLF, if present, would require redshift bins finer than the current detected sample size
supports without the Schechter fits becoming prior-dominated (Sect.~\ref{sec:xlf_robustness}),
together with a per-group AGN classification of the same sample and, ideally, a matched comparison
against cosmological simulations with AGN feedback (e.g.\ IllustrisTNG, EAGLE, SIMBA, Magneticum).
We defer that joint analysis to future work.

Details of the public data release, including the full column dictionary and the highest-significance
individual CW-HCG detections, are given in Appendix~\ref{app:release}.

\section{Summary}
\label{sec:conclusions}

We have presented the X-ray characterization of the COSMOS-Web galaxy group population: 1,678
groups identified with AMICO \citep[CW-All;][]{Toni2025} and 993 Hickson compact groups (CW-HCG;
Hasinger et al., in preparation) identified independently within the same
$0.45~\mathrm{deg}^2$ footprint, spanning $0.08 \lesssim z \lesssim 3.8$. Group spectroscopic
membership and BGG identification (Sect.~\ref{sec:membership}) are used to set the X-ray aperture
centering and geometry, and a uniform aperture-photometry X-ray pipeline is applied to every group
using the combined Chandra and XMM-Newton COSMOS mosaic (Sect.~\ref{sec:analysis}). We detect
X-ray emission from 499 CW-All groups (29.7\%) and 332 CW-HCG groups (33.4\%) at SNR~$\geq 2.0$
(Sect.~\ref{sec:catalog_overview}); ensemble properties of the non-detected majority, recovered
via X-ray stacking, are presented in Paper~II. The X-ray luminosity function of the detected
population, presented in four redshift bins spanning $0.05 \leq z < 3.80$ as the principal
statistical result of this paper (Sect.~\ref{sec:xlf}), is consistent with a steep, negatively
evolving group-scale XLF whose characteristic luminosity $L^\star$ rises by $\sim$1.9--2.1~dex
from the lowest to the highest redshift bin in both catalogs. This rise is statistically robust in
the upper half of our redshift range (beyond 95\% confidence for the steepest step in each
catalog; Appendix~\ref{app:xlf_robustness}) and is not explained by mass or richness selection of
the detected population, as traced by four independent, X-ray-independent proxies ($\lambda_\star$,
membership richness, discovery richness, and BGG stellar mass; Sect.~\ref{sec:selection_effects}).
A targeted, field-calibrated search for diffuse X-ray emission from the AmicOne protocluster
candidate \citep{Toni2025} -- combining a tile-based blind search on the source-masked map with an
independent test against the projected galaxy overdensity, which avoids relying on the AmicOne
group membership itself -- finds no statistically significant signal at the current depth
(whole-box $3\sigma$ upper limit $\log_{10}(L_{\rm X}/\mathrm{erg~s}^{-1}) < 44.12$), though
fainter, patchy emission below our sensitivity cannot be excluded (Appendix~\ref{app:amicone}).

We publicly release the X-ray fluxes, luminosities, and derived properties for both samples
(Appendix~\ref{app:release}) as reference data products for this survey; the parent optical/NIR
group catalogs themselves are released in \citet{Toni2025} for CW-All and will be released in
Hasinger et al. (in preparation) for CW-HCG. These X-ray measurements, together with the pipeline
and quality flags documented here, are intended to serve as the common basis for the companion and
future papers in this series: Paper~II presents the full CW-HCG X-ray and stacking analysis and a
direct comparison with CW-All; the halo masses reported here already adopt the weak-lensing
calibrated $M$--$L_{\rm X}$ relation of the companion paper \citet{Leroy2026} (in prep.), which uses JWST
weak-lensing halo masses measured directly in the COSMOS-Web footprint to supersede
\citet{Leauthaud2010}; and subsequent papers will address AGN demographics,
stellar-mass--$L_{\rm X}$ scaling, dynamical mass calibration, environmental dependence, and the
conversion of the XLF presented here into a group-scale halo mass function via the same $M$--$L_{\rm
X}$ relation, using the X-ray measurements released here. Together, the uniform X-ray pipeline, the
quality flags, and
the public X-ray measurements presented here establish the common empirical foundation on which
this series of papers builds.

Looking ahead, planned and proposed X-ray facilities would directly address limitations of the
present analysis. ESA's Athena mission, and in particular its X-IFU microcalorimeter, would enable
direct spectroscopic temperature measurements for individual COSMOS-Web groups that are presently
characterized only through the scaling relations of Sect.~\ref{sec:scaling}. Proposed probe-class
missions such as AXIS, combining sub-arcsecond angular resolution with deep sensitivity, would
resolve individual faint, high-redshift groups that our current depth can only characterize in
aggregate through stacking (Paper~II). Combined with the JWST-based spectroscopic and structural
information already available in COSMOS-Web, such facilities would substantially sharpen the
group-scale picture this series of papers is beginning to establish.

\begin{acknowledgements}
The authors acknowledge support from the Academy of Finland, the Jane and Aatos Erkko
Foundation, the Finnish Ministry of Education and Culture (Global Programme), and the EuroHPC
Plasma-AI project (RCF project number 373416).
\end{acknowledgements}


\begin{appendix}
\section{Public data release}
\label{app:release}
The X-ray measurements described in this paper are available at
\url{https://github.com/gozaliasl/cosmos-web-xray-catalog-release}. The parent optical/NIR group
catalogs on which this X-ray analysis is based are released separately, in \citet{Toni2025} for
CW-All and in Hasinger et al. (in preparation) for CW-HCG.

We publicly release the X-ray measurements and derived properties for the individually detected
(SNR~$\geq 2.0$) AMICO-selected and Hickson compact-group samples as reference data products of
this survey, as a group-level FITS/CSV table for each catalog. \texttt{ID} is \texttt{Group\_ID},
not assigned internally by this
pipeline but inherited unchanged from the parent catalogs (from the AMICO group catalog of
\citet{Toni2025} for CW-All, and from the Hickson compact-group catalog of Hasinger et al.\ in
preparation for CW-HCG; Sect.~\ref{sec:hcg_hasinger}), so that users cross-matching to either parent
catalog, or to companion papers in this series, can do so directly on \texttt{Group\_ID} without an
additional positional match; \texttt{RA}, \texttt{Dec}, and \texttt{Redshift} are the group
centroid coordinates and the redshift used for the X-ray analysis; \texttt{SNR} is the
aperture-photometry detection significance; \texttt{Flux\_erg\_cm2\_s} and \texttt{Luminosity\_erg\_s}
are the observed-frame 0.5--2~keV flux and rest-frame, K-corrected 0.5--2~keV luminosity
(Sect.~\ref{sec:analysis}); \texttt{Temperature\_keV} is the spectral-model temperature
(Sect.~\ref{sec:scaling}); \texttt{Log10\_M200\_Luminosity} is the weak-lensing calibrated,
luminosity-based halo mass from the \citet{Leroy2026} (in prep.) $M$--$L_{\rm X}$ scaling relation,
the primary mass estimate adopted throughout this paper, with \texttt{R500\_Luminosity\_kpc} and
\texttt{R200\_Luminosity\_kpc} the corresponding overdensity radii; and
\texttt{Is\_Projected\_Contaminated} and \texttt{Is\_Suspected\_False\_Positive} are the projected-
contamination and suspected-false-positive quality flags. All quantities carry their $1\sigma$
uncertainties where applicable. This release contains X-ray measurements for individually detected
groups only; non-detections/upper limits, background diagnostics, binned/stacking-analysis columns,
and the temperature-based mass alternative are not included here. Group membership and richness are
part of the parent group catalogs themselves \citep[][Hasinger et al., in preparation, for
CW-HCG]{Toni2025}, and BGG identification is presented in \citet{Gozaliasl2025}; cross-group
dynamical-mass and multiplicity information are deferred to a separate, future companion release.
The X-ray measurements presented here incorporate a corrected energy-conversion
factor, an updated Planck 2018 cosmology rescaling \citep{Aghanim2020}, and an improved
background-quality flag relative to an earlier internal version of this catalog. Companion papers
building on this release will use identical group and galaxy IDs, so that results across the series
can be directly cross-matched.

Table~\ref{tab:xray_sample_hcg} reproduces the released columns in full for the 20
highest-significance CW-HCG detections.

\begin{table*}[t]
\caption{X-ray properties of the 20 highest-significance individually detected groups in the CW-HCG sample (SNR~$\geq
  10.4$), sorted by detection significance. Column definitions are identical to Table~\ref{tab:detection}. Flag: C
  indicates \texttt{Is\_Projected\_Contaminated}; F indicates \texttt{Is\_Suspected\_False\_Positive}.}
\label{tab:xray_sample_hcg}
\centering
\setlength{\tabcolsep}{4pt}
\begin{tabular}{rcccrcccrrrc}
\hline\hline
ID & RA & Dec & $z$ & SNR & $f_{\rm X}$ & $\log L_{\rm X}$ & $T_{\rm X}$ & $\log M_{200}$ & $R_{500}$ & $R_{200}$ & Flag \\
  & (deg) & (deg) & & & ($10^{-14}$ c.g.s.) & (erg~s$^{-1}$) & (keV) & ($M_\odot$) & (kpc) & (kpc) & \\
\hline
20975 & 150.4472 & 2.4195 & 0.215 & 28.7 & $0.59\pm0.05$ & $42.37\pm0.02$ & $0.82\pm0.03$ & $13.58\pm0.22$ & 434 & 659 & -- \\
25137 & 150.4257 & 2.4470 & 0.388 & 28.6 & $5.93\pm0.11$ & $42.93\pm0.02$ & $2.74\pm0.02$ & $13.92\pm0.16$ & 528 & 802 & -- \\
26731 & 150.4211 & 2.4330 & 2.451 & 27.3 & $6.54\pm0.15$ & $45.35\pm0.02$ & $10.14\pm0.09$ & $14.92\pm0.45$ & 550 & 835 & C \\
102675 & 150.4223 & 2.4305 & 1.519 & 27.1 & $6.89\pm0.15$ & $44.69\pm0.02$ & $7.68\pm0.07$ & $14.75\pm0.35$ & 645 & 980 & C \\
7983  & 149.9228 & 2.5253 & 0.731 & 26.3 & $4.47\pm0.11$ & $43.72\pm0.02$ & $4.00\pm0.04$ & $14.36\pm0.20$ & 645 & 980 & -- \\
10778 & 150.4132 & 2.4090 & 0.385 & 25.6 & $1.36\pm0.06$ & $42.81\pm0.02$ & $1.68\pm0.03$ & $13.83\pm0.17$ & 494 & 750 & -- \\
7998  & 149.9142 & 2.5294 & 0.733 & 21.9 & $4.23\pm0.11$ & $43.53\pm0.02$ & $3.94\pm0.04$ & $14.22\pm0.17$ & 579 & 878 & -- \\
55942 & 150.4381 & 2.4304 & 2.326 & 20.4 & $4.12\pm0.12$ & $44.98\pm0.02$ & $8.43\pm0.10$ & $14.68\pm0.37$ & 473 & 718 & C \\
65991 & 150.4316 & 2.4081 & 1.629 & 15.3 & $1.67\pm0.07$ & $44.25\pm0.03$ & $5.02\pm0.09$ & $14.37\pm0.25$ & 464 & 705 & C \\
23848 & 150.4106 & 2.4498 & 0.816 & 15.0 & $2.53\pm0.08$ & $43.34\pm0.03$ & $3.59\pm0.05$ & $14.03\pm0.16$ & 484 & 735 & -- \\
18746 & 149.9234 & 2.5384 & 0.550 & 13.8 & $3.49\pm0.09$ & $42.85\pm0.03$ & $2.99\pm0.03$ & $13.78\pm0.17$ & 446 & 677 & -- \\
156907 & 150.1879 & 1.7795 & 0.764 & 12.1 & $0.74\pm0.05$ & $43.07\pm0.04$ & $2.28\pm0.06$ & $13.86\pm0.16$ & 431 & 655 & -- \\
32973 & 150.1125 & 2.3857 & 0.263 & 11.4 & $0.51\pm0.04$ & $41.89\pm0.04$ & $0.91\pm0.03$ & $13.19\pm0.32$ & 317 & 481 & -- \\
21366 & 150.0675 & 2.4232 & 0.083 & 11.1 & $0.15\pm0.03$ & $40.83\pm0.04$ & $0.27\pm0.02$ & $12.46\pm0.54$ & 193 & 293 & -- \\
10141 & 150.4023 & 2.4641 & 0.503 & 11.0 & $0.66\pm0.04$ & $42.54\pm0.04$ & $1.61\pm0.04$ & $13.57\pm0.21$ & 386 & 586 & -- \\
20904 & 150.3966 & 2.5172 & 0.219 & 10.7 & $0.35\pm0.03$ & $41.68\pm0.04$ & $0.70\pm0.03$ & $13.05\pm0.36$ & 289 & 439 & -- \\
18480 & 150.3920 & 2.4336 & 0.615 & 10.7 & $1.01\pm0.05$ & $42.74\pm0.04$ & $2.15\pm0.04$ & $13.68\pm0.18$ & 399 & 606 & -- \\
12520 & 150.1132 & 2.3557 & 0.220 & 10.6 & $0.41\pm0.04$ & $41.79\pm0.04$ & $0.75\pm0.03$ & $13.13\pm0.34$ & 307 & 467 & -- \\
12400 & 150.1221 & 2.3504 & 0.222 & 10.6 & $0.24\pm0.04$ & $41.79\pm0.04$ & $0.63\pm0.04$ & $13.13\pm0.34$ & 307 & 466 & -- \\
3143  & 149.9248 & 2.5010 & 1.426 & 10.4 & $0.95\pm0.05$ & $43.84\pm0.04$ & $3.82\pm0.09$ & $14.15\pm0.19$ & 418 & 635 & -- \\
\hline
\multicolumn{12}{l}{\textit{Note.} The full catalog of all 332 CW-HCG detections is available in machine-readable form.}
\end{tabular}
\end{table*}

\begin{table*}[t]
\caption{X-ray properties of the 20 highest-significance individually detected groups in the CW-All sample, sorted by
  detection significance. Column definitions are identical to Table~\ref{tab:detection}. Flag: C indicates
  \texttt{Is\_Projected\_Contaminated}; F indicates \texttt{Is\_Suspected\_False\_Positive}.}
\label{tab:xray_sample_all}
\centering
\setlength{\tabcolsep}{4pt}
\begin{tabular}{rcccrcccrrrc}
\hline\hline
ID & RA & Dec & $z$ & SNR & $f_{\rm X}$ & $\log L_{\rm X}$ & $T_{\rm X}$ & $\log M_{200}$ & $R_{500}$ & $R_{200}$ & Flag \\
  & (deg) & (deg) & & & ($10^{-14}$ c.g.s.) & (erg~s$^{-1}$) & (keV) & ($M_\odot$) & (kpc) & (kpc) & \\
\hline
41 & 150.4441 & 2.4172 & 0.155 & 39.3 & $1.72\pm0.06$ & $42.31\pm0.01$ & $0.93\pm0.01$ & $13.47\pm0.01$ & 410 & 622 & -- \\
376 & 150.4191 & 2.4572 & 0.216 & 28.7 & $5.26\pm0.10$ & $42.34\pm0.02$ & $1.71\pm0.01$ & $13.46\pm0.01$ & 397 & 603 & -- \\
1 & 149.9186 & 2.5172 & 0.711 & 28.2 & $4.61\pm0.12$ & $43.72\pm0.02$ & $3.96\pm0.04$ & $14.14\pm0.01$ & 548 & 833 & -- \\
1122 & 150.4891 & 2.2372 & 0.193 & 18.1 & $0.27\pm0.04$ & $41.97\pm0.02$ & $0.59\pm0.03$ & $13.24\pm0.02$ & 338 & 513 & -- \\
1395 & 150.4241 & 2.4472 & 3.020 & 17.8 & $2.11\pm0.08$ & $45.25\pm0.02$ & $7.85\pm0.12$ & $14.32\pm0.02$ & 299 & 454 & C \\
237 & 149.9236 & 2.5072 & 2.584 & 17.5 & $1.98\pm0.09$ & $45.05\pm0.02$ & $7.04\pm0.12$ & $14.31\pm0.02$ & 332 & 503 & C \\
10 & 150.1688 & 1.7822 & 0.167 & 16.9 & $0.81\pm0.05$ & $41.78\pm0.03$ & $0.76\pm0.02$ & $13.13\pm0.02$ & 313 & 475 & -- \\
58 & 150.1839 & 1.7772 & 0.473 & 16.1 & $1.13\pm0.06$ & $42.75\pm0.03$ & $1.84\pm0.04$ & $13.62\pm0.02$ & 405 & 615 & -- \\
868 & 150.4191 & 2.4072 & 1.272 & 15.8 & $2.12\pm0.07$ & $43.90\pm0.03$ & $4.62\pm0.06$ & $14.02\pm0.02$ & 401 & 609 & -- \\
15 & 150.0788 & 2.3922 & 0.220 & 14.3 & $1.05\pm0.06$ & $41.91\pm0.03$ & $1.02\pm0.02$ & $13.19\pm0.02$ & 322 & 489 & -- \\
330 & 150.1688 & 1.7822 & 0.278 & 12.8 & $0.93\pm0.05$ & $42.04\pm0.03$ & $1.16\pm0.03$ & $13.25\pm0.02$ & 330 & 501 & -- \\
124 & 150.3891 & 2.4172 & 0.326 & 12.4 & $1.20\pm0.06$ & $42.15\pm0.04$ & $1.42\pm0.03$ & $13.30\pm0.02$ & 337 & 511 & -- \\
78 & 150.3640 & 2.4672 & 0.118 & 11.3 & $0.21\pm0.03$ & $41.26\pm0.04$ & $0.38\pm0.02$ & $12.82\pm0.02$ & 251 & 380 & -- \\
1739 & 150.1789 & 1.7772 & 0.970 & 11.3 & $0.96\pm0.05$ & $43.38\pm0.04$ & $2.95\pm0.07$ & $13.81\pm0.02$ & 383 & 581 & -- \\
17 & 150.1238 & 2.3522 & 0.210 & 11.0 & $0.39\pm0.05$ & $41.75\pm0.04$ & $0.71\pm0.03$ & $13.09\pm0.03$ & 300 & 455 & -- \\
524 & 150.4041 & 2.4622 & 0.516 & 10.7 & $1.95\pm0.07$ & $42.55\pm0.04$ & $2.35\pm0.03$ & $13.47\pm0.03$ & 356 & 540 & -- \\
234 & 149.9386 & 2.5872 & 0.079 & 10.6 & $0.06\pm0.02$ & $40.72\pm0.04$ & $0.19\pm0.03$ & $12.48\pm0.03$ & 196 & 298 & -- \\
600 & 150.1038 & 2.3672 & 0.329 & 10.6 & $0.26\pm0.04$ & $42.08\pm0.04$ & $0.87\pm0.05$ & $13.25\pm0.03$ & 324 & 491 & -- \\
1737 & 149.9337 & 2.5022 & 3.280 & 10.3 & $0.67\pm0.05$ & $45.01\pm0.04$ & $5.63\pm0.17$ & $14.10\pm0.03$ & 238 & 362 & C \\
961 & 149.9336 & 2.6072 & 0.800 & 9.8 & $0.50\pm0.04$ & $43.00\pm0.04$ & $2.06\pm0.07$ & $13.64\pm0.03$ & 360 & 547 & -- \\
\hline
\multicolumn{12}{l}{\textit{Note.} The full catalog of all 499 CW-All detections is available in machine-readable form.}
\end{tabular}
\end{table*}

\section{Detection and fit diagnostics}
\label{app:diagnostics}

\subsection{Upper-limit diagnostics}
\label{app:upper_limits_fig}

\begin{figure*}[htbp]
\centering
\includegraphics[width=0.7\textwidth]{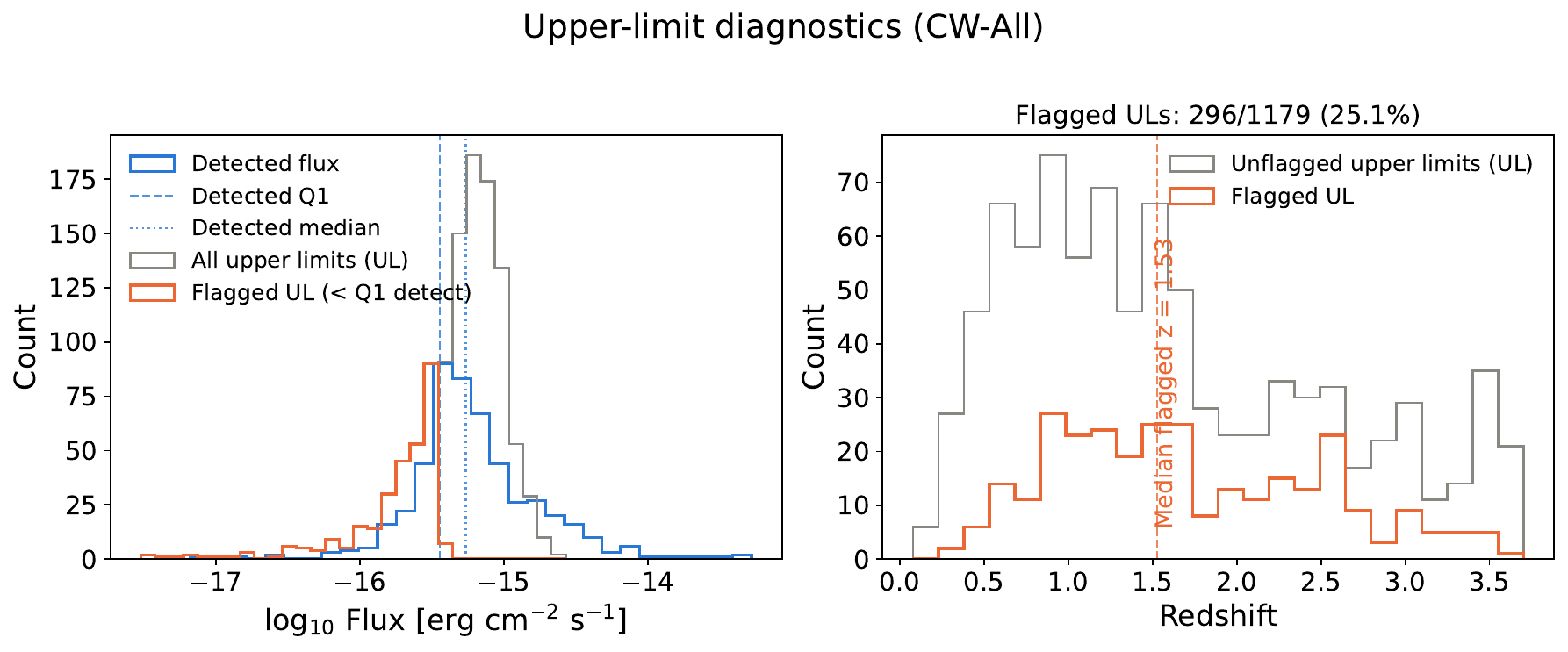}\\[4pt]
\includegraphics[width=0.7\textwidth]{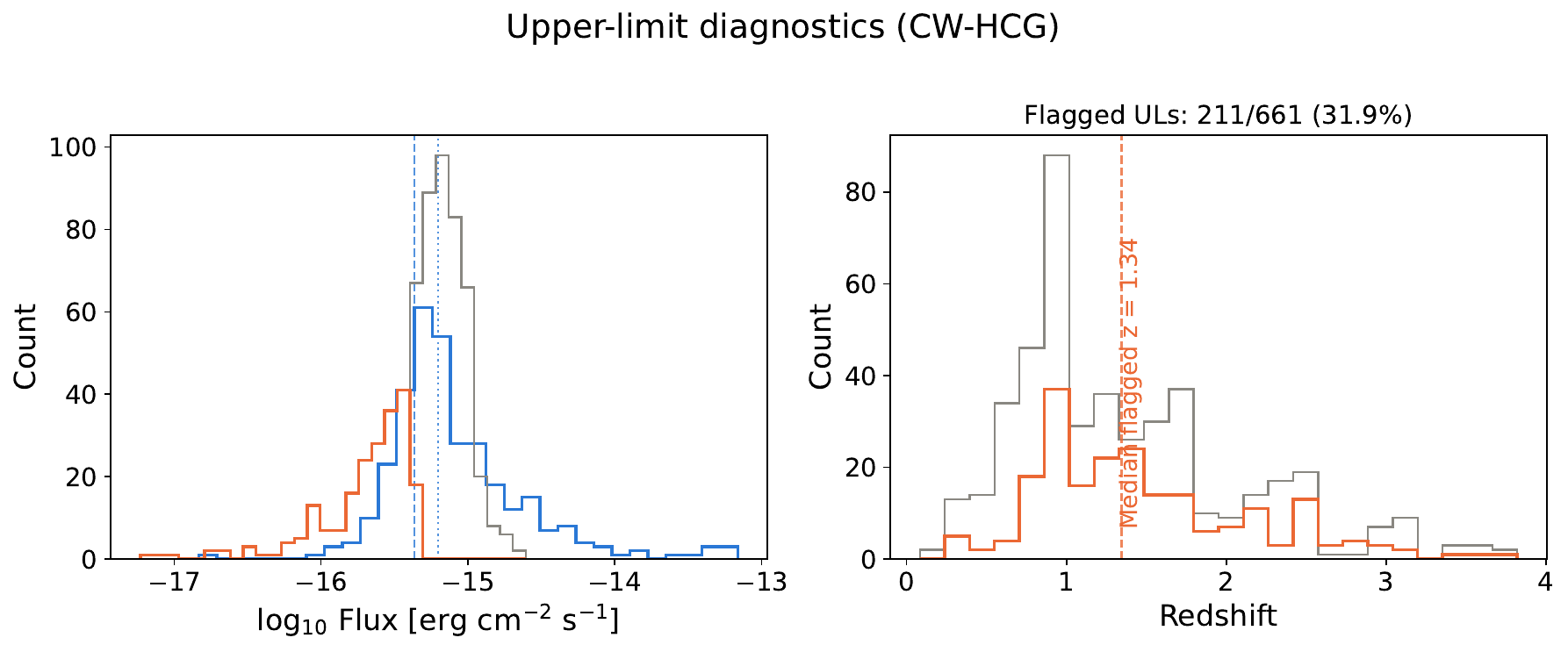}
\caption{Upper-limit (UL) diagnostics for CW-All (top) and CW-HCG (bottom). Left panels: flux
distribution histogram showing detected sources (blue), all ULs (gray, all non-detections), and
flagged ULs (orange, the subset of ULs that fall below the first quartile of the detected flux
distribution); both UL histograms share identical bin edges, so the orange curve is a true subset
of the gray one in every bin. Vertical dashed
and dotted lines mark the first quartile (Q1) and median of detected fluxes. Right panels: redshift
distribution of the flagged ULs versus the remaining, unflagged ULs (i.e.\ all non-detections split
into these two mutually exclusive subsets). Flagged sources are monitored for quality control;
masking them changes the stacked fluxes by less than the statistical uncertainties.}
\label{fig:upper_limits}
\end{figure*}

Fig.~\ref{fig:upper_limits} shows the full upper-limit diagnostic plots for both catalogs,
underlying the summary given in Sect.~\ref{sec:catalog_overview}.

\subsection{XLF Schechter-fit posterior example}
\label{app:corner}

\begin{figure}[htbp]
\centering
\includegraphics[width=\columnwidth]{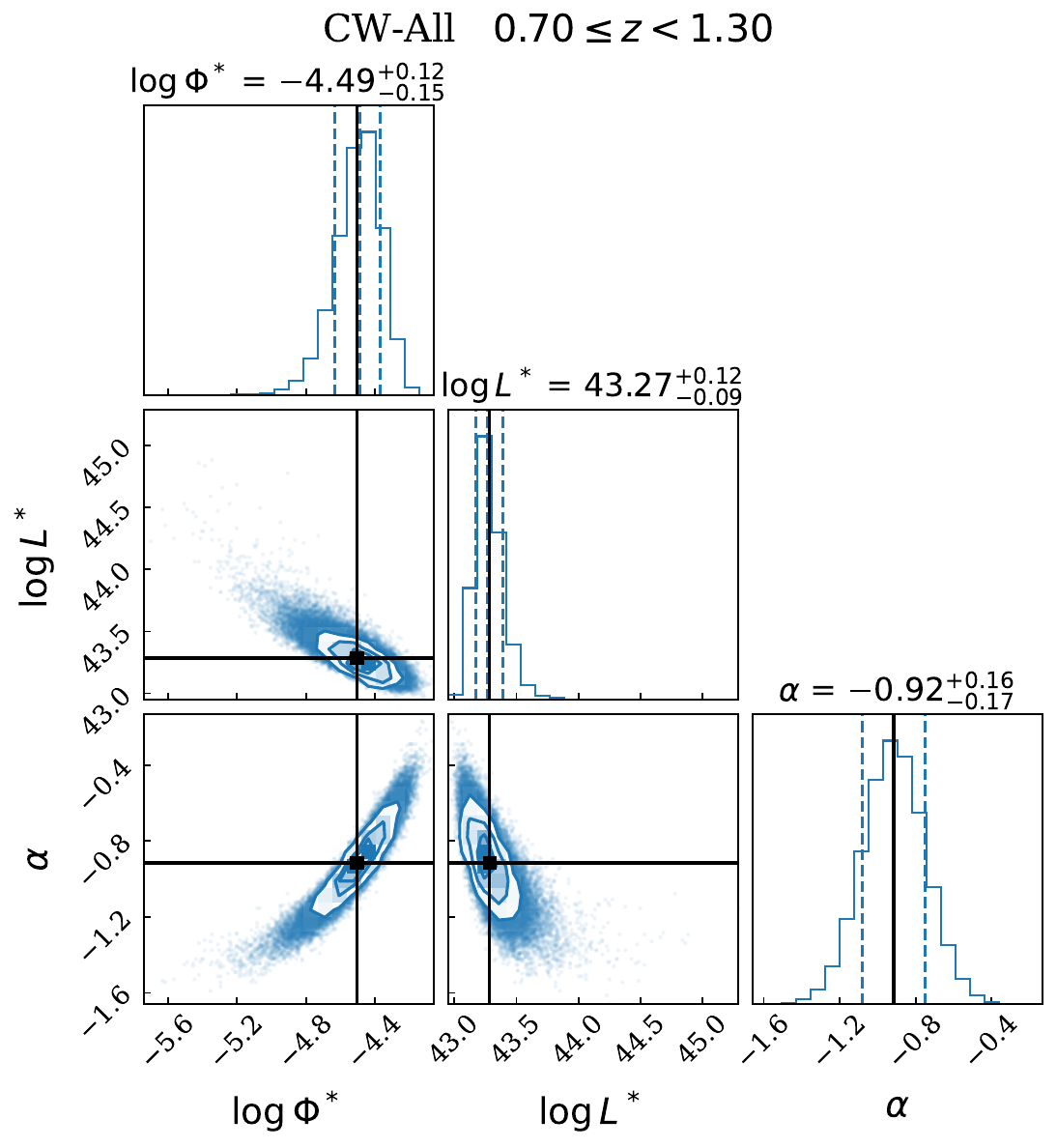}
\caption{Posterior distributions and credible regions for the Schechter fit parameters
($\log\Phi^\star$, $\log L^\star$, $\alpha$) of the CW-All XLF at $0.70 \leq z < 1.30$
(Table~\ref{tab:xlf_schechter}, row 3), from 128-walker, 12,000-step MCMC sampling with a 3,000-step
burn-in; this bin is shown as a representative well-constrained fit, with tighter and less
asymmetric credible intervals than the lower-$z$ bin.}
\label{fig:xlf_corner}
\end{figure}

\section{A compact-group showcase at $z\sim2.5$: CW-HCG~01101}
\label{app:showcase}

\begin{figure}[htbp]
\centering
\includegraphics[width=0.9\columnwidth]{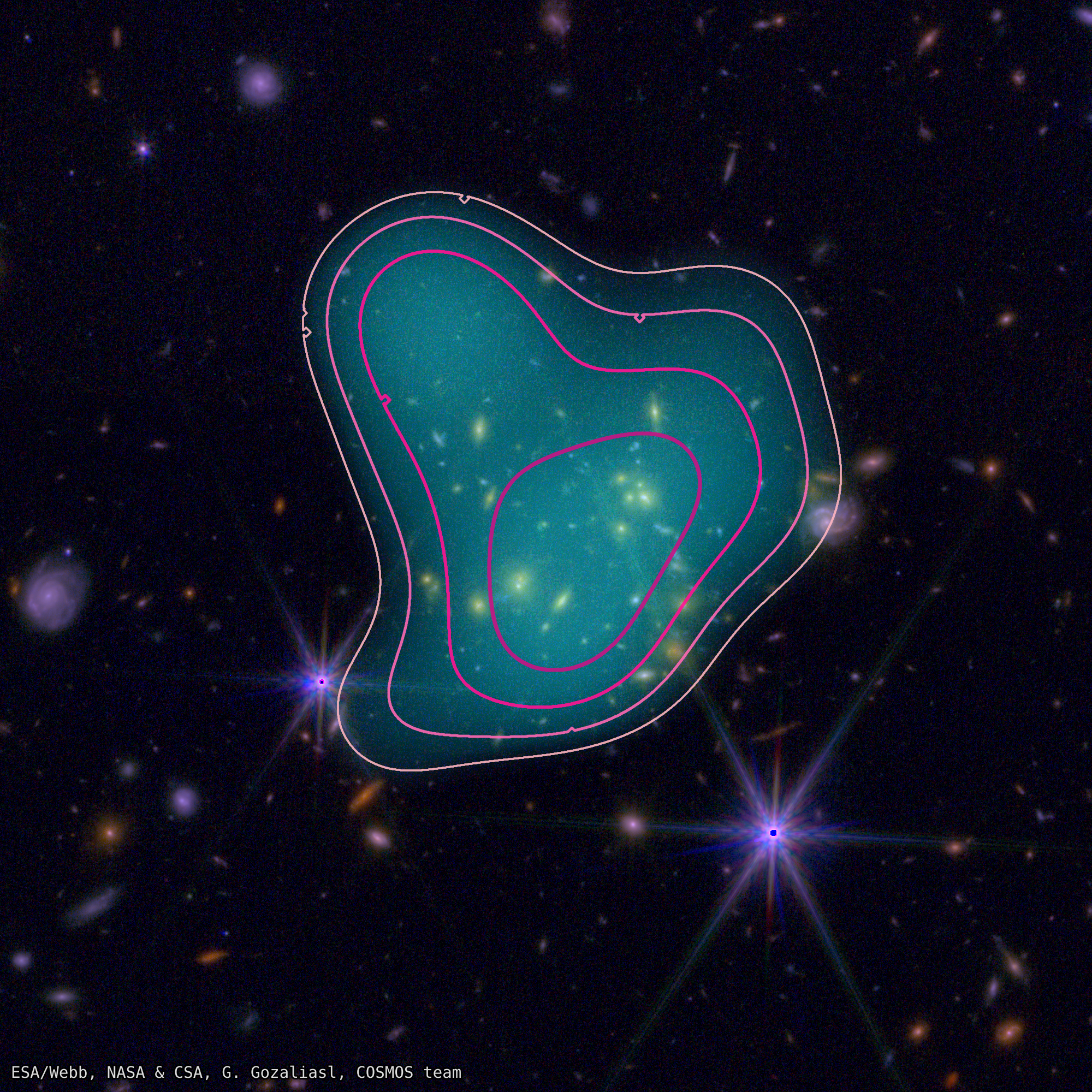}
\caption{JWST/COSMOS-Web RGB cutout of CW-HCG~01101 at $z=2.508$, overlaid with X-ray significance contours (Chandra+XMM-Newton, Gaussian-smoothed), drawn at the 70th and 90th percentiles of the smoothed count-rate distribution (or, where these are not distinguishable, at the median and peak, approximating the half-maximum and peak extent of the diffuse emission). The group is individually detected at SNR~$=3.4$ with $\log_{10}(L_{\rm X}/\mathrm{erg~s}^{-1})=43.8$ and $\log_{10}(M_{200}/M_\odot)=13.5$, and lies in the same region of COSMOS as the distant X-ray cluster reported by \citet{Wang2016}.}
\label{fig:group01070}
\end{figure}

As an illustration of an individual detection in the high-redshift regime discussed in
Sect.~\ref{sec:showcase}, Fig.~\ref{fig:group01070} shows CW-HCG group 01101, a Hickson-style
compact group at $z=2.508$
($\mathrm{RA}=150.2386^{\circ}$, $\mathrm{Dec}=2.3359^{\circ}$). The system is individually X-ray
detected at SNR~$=3.4$ ($f_{\rm X} = 5.4\times10^{-16}$~erg~cm$^{-2}$~s$^{-1}$,
$\log_{10}(L_{\rm X}/\mathrm{erg~s}^{-1}) = 43.8$), yielding $T_{\rm X}=2.1$~keV and
$\log_{10}(M_{200}/M_\odot)=13.5$ ($R_{200}=278$~kpc) from the temperature-based
scaling relation; it is flagged neither as projection-contaminated nor as a suspected false
positive. It lies below the SNR~$\geq10.4$ threshold of Table~\ref{tab:xray_sample_hcg}, so it is
not listed there explicitly, but among the 18 clean CW-HCG detections at $z\geq2.0$ identified in
Sect.~\ref{sec:showcase}, it is the system closest to the redshift and diffuse-emission morphology
reported for a
known distant X-ray-detected cluster hosting a starbursting core in the same region of the COSMOS
field \citep{Wang2016}. This system has no corresponding CW-All counterpart: the nearest CW-All
group lies $17\arcsec$ away but at a significantly different redshift ($z=2.37$), so it is a
distinct system rather than the same structure recovered by both group finders. The CW-HCG
detection presented here is derived independently, using
COSMOS-Web/JWST-based group selection and membership together with the Chandra+XMM-Newton pipeline
described in Sect.~\ref{sec:analysis}, and is consistent with an extended, group-scale hot halo
already in place at $z\sim2.5$. Together with the compact-group selection discussed in
Sect.~\ref{sec:hcg_hasinger}, this system exemplifies the value of combining JWST-depth
compact-group identification with sensitive X-ray follow-up to probe the assembly of hot gaseous
halos at the earliest cosmic epochs accessible to this survey.

\section{Background method comparison}
\label{app:background}

To quantify the impact of the background-estimation scheme, we compared photometry obtained with the
default per-source annulus background to measurements using the redshift-binned median background.
For the CW-All catalog, the median ratio of binned-median to annulus-based $L_{\rm X}$ is 1.005 with
a scatter of 0.135~dex, while the corresponding flux ratio is 0.992 with a scatter of 0.195~dex.
These values indicate that the two background schemes are essentially unbiased with respect to each
other, and that the additional systematic scatter introduced by the choice of background is at the
$\sim$30--60\% level, sub-dominant to the statistical uncertainties for most individual groups. We
therefore adopt the annulus-based measurements for our main analysis.

\begin{figure*}[htbp]
\centering
\includegraphics[width=0.7\textwidth]{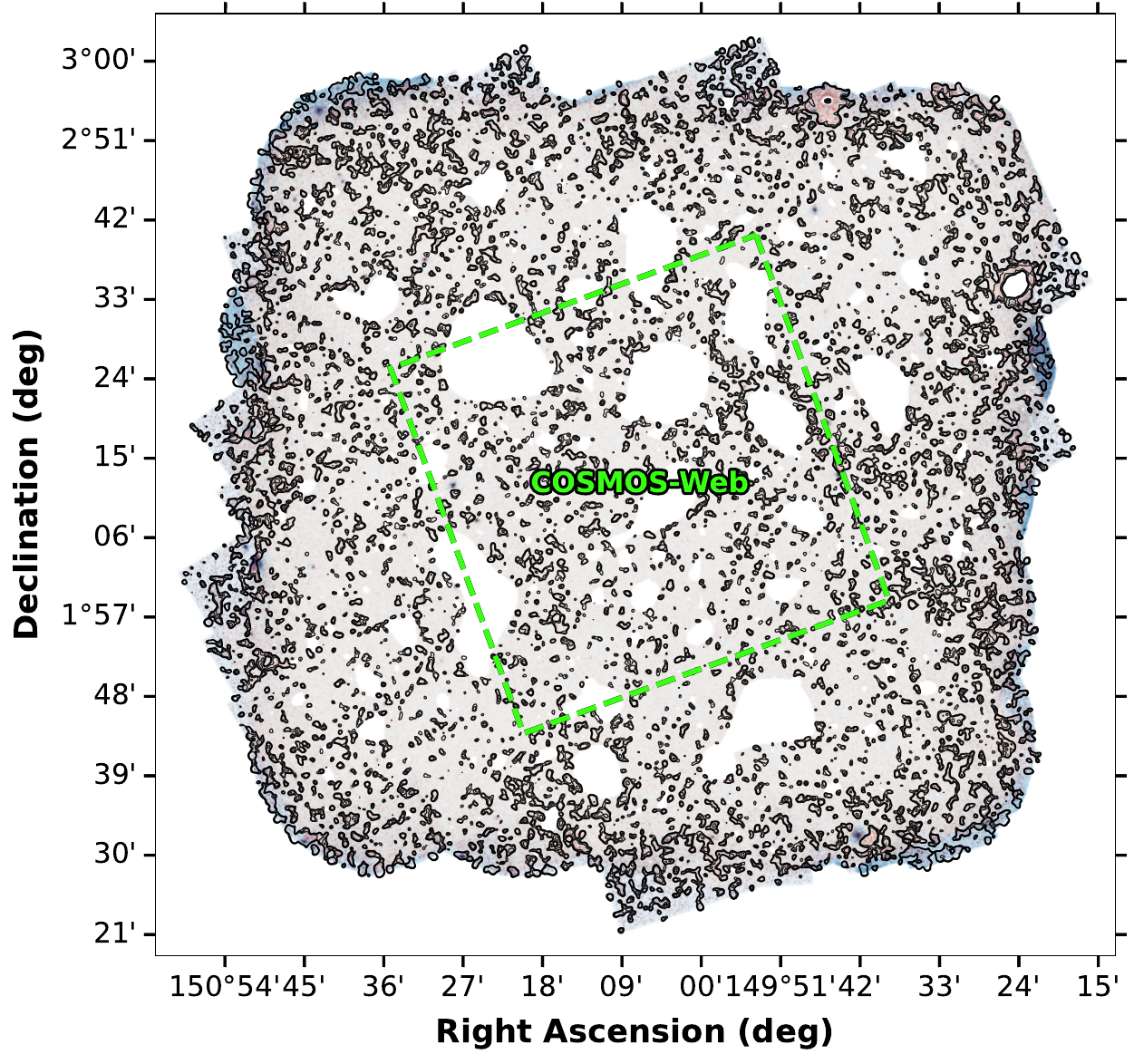}
\caption{The masked (source-excised) version of the full COSMOS X-ray map (Fig.~\ref{fig:xray_map}),
used for background determination and X-ray characterization of groups at $z\geq1.5$
(Sect.~\ref{sec:xraydata}), and for the diffuse, stacking-oriented analyses of this series
(Paper~II). White regions mark all detected X-ray sources, masked out to $2\,R_{200}$ of each
detected group, in addition to point sources, leaving only clean, diffuse background emission. The
lime dashed polygon indicates the COSMOS-Web footprint.}
\label{fig:xray_map_masked}
\end{figure*}

\begin{figure}[htbp]
\centering
\includegraphics[width=\linewidth]{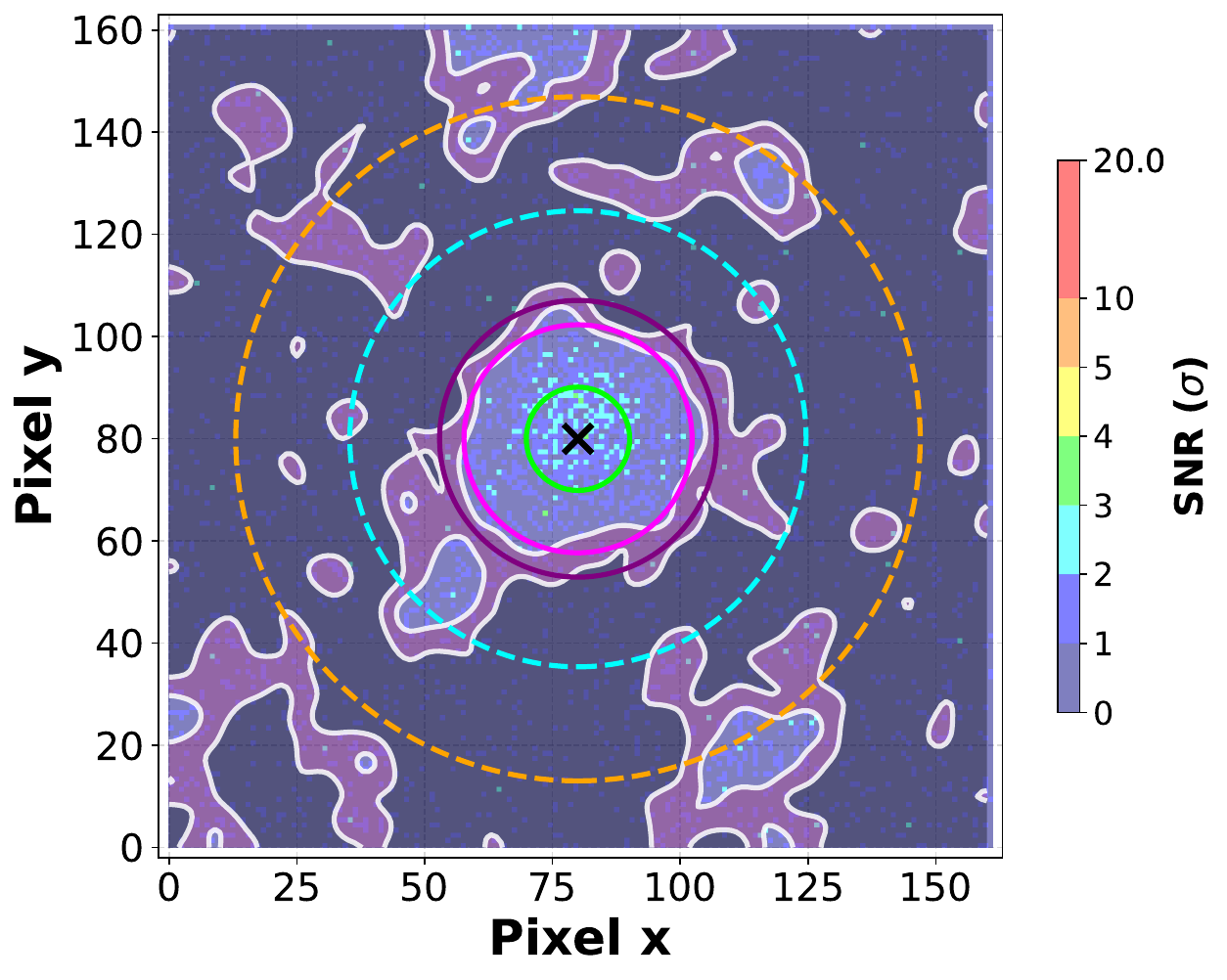}
\caption{X-ray significance (SNR) map of Group~1 with the pipeline's aperture and background
geometry overlaid (Sect.~\ref{sec:analysis}). Green: 300~kpc first-pass source aperture used for
detection (Sect.~\ref{sec:aperture}). Magenta/dark-red: the second-pass, mass-derived
$R_{500}$/$R_{200}$ aperture used to remeasure detected sources. Cyan/orange dashed: the adaptive
background annulus ($2$--$3\,R_{500}$). Black cross: the group's catalog (optical/near-infrared)
center.}
\label{fig:group1_snr_appendix}
\end{figure}

Fig.~\ref{fig:group1_snr_appendix} shows the full aperture and background geometry
(Sect.~\ref{sec:analysis}) underlying the Group~1 showcase of Fig.~\ref{fig:group1_showcase}.

\section{No significant diffuse X-ray emission from the AmicOne protocluster wall}
\label{app:amicone}

As a further test of the high-$z$ population, we searched for diffuse, structure-scale X-ray
emission associated with ``AmicOne,'' the largest protocluster candidate identified by
\citet{Toni2025} from a three-dimensional clustering analysis of AMICO detections: 14 density-peak
cores at $2.5 \leq z \leq 3.09$ spanning $\sim$20~cMpc on the sky and $\sim$600~cMpc in the radial
direction, within a $150.175\degr$--$150.375\degr$ (RA) by $1.750\degr$--$1.950\degr$ (Dec) box.
This $\sim$600~cMpc line-of-sight extent is far larger than any single gravitationally bound
structure could span, and more plausibly reflects the photometric-redshift uncertainty of the
individual AMICO group detections that make up the 14 cores than a protocluster progenitor observed
coherently at one cosmic epoch; we nonetheless search the full box as defined by \citet{Toni2025},
since narrowing it ourselves would be an arbitrary, unvalidated choice. Fifteen CW-All groups fall
within this footprint and redshift range -- a superset of, not identical to, the 14 formally
identified density-peak cores, since our search box also contains other AMICO groups that lie in
the same region without individually registering as one of the 14 peaks -- of which only one is
individually flagged as detected (SNR$=2.5$), and that detection is itself flagged as
projection-contaminated. A catalog-level ensemble stack of all 15 apertures gives SNR$\,\approx\,1.4$,
consistent with no signal.

\begin{figure*}[htbp]
\centering
\includegraphics[width=0.95\textwidth]{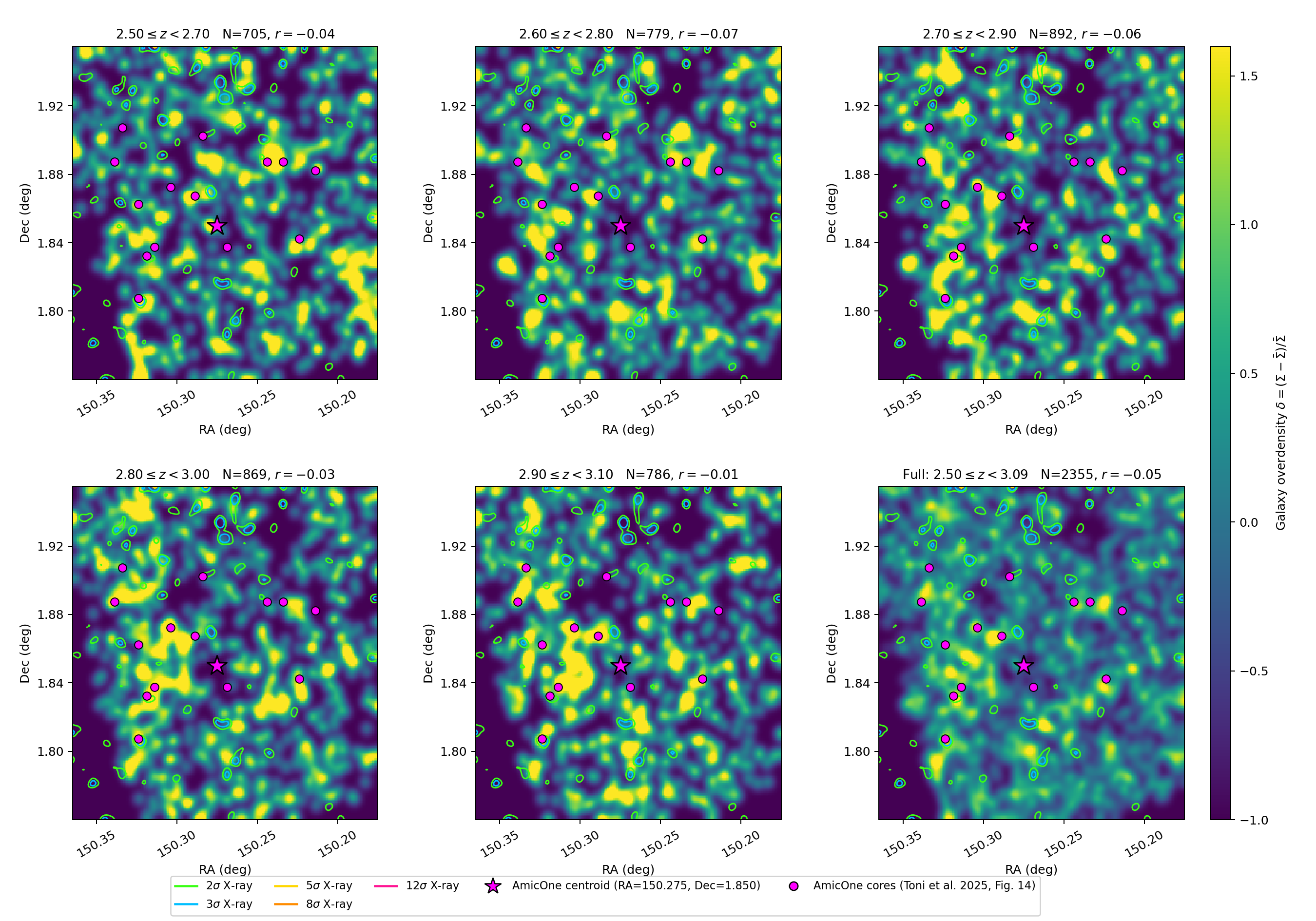}
\caption{Projected galaxy overdensity $\delta=(\Sigma-\bar\Sigma)/\bar\Sigma$ in the AmicOne box,
computed independently of any group catalog from the full COSMOS-Web photometric sample
(\texttt{LP\_type}$=0$, \texttt{MAG\_MODEL\_F444W}$<27.5$), for five sliding $\Delta z=0.20$ redshift
sub-slices spanning the AmicOne window plus the full $2.50\leq z<3.09$ slice (bottom right). Contours
mark the source-masked X-ray significance at 2, 3, 5, 8, and 12$\sigma$ (green through pink), computed
from the same map used in the tile-based null test above. $N$ is the number of selected galaxies in
each slice and $r$ is the Pearson cross-correlation coefficient between the overdensity and X-ray
significance maps. No slice shows a positive or preferred alignment between the galaxy density peaks
and the X-ray patches.}
\label{fig:amicone_null}
\end{figure*}

We then tested for genuinely diffuse (rather than per-group) emission using the source-masked
X-ray map, from which small-scale (point-like) emission has already been removed by construction,
independently of any external source catalog. We consider two complementary ways of integrating
this map over the full AmicOne box. First, summing only the pixels that individually exceed a
$2\sigma$ (respectively $3\sigma$) significance threshold -- i.e.\ taking the $2\sigma$ patches
visible in Fig.~\ref{fig:amicone_null} at face value -- gives an integrated count rate of
$2.40\times10^{-3}$ ($8.86\times10^{-4}$)~counts~s$^{-1}$ over 1062 (234) pixels, corresponding to
$\log_{10}(L_{\rm X}/\mathrm{erg~s}^{-1}) = 44.33$ (43.90) at $z=2.65$. This threshold-selected sum
is, by construction, a positively biased estimator: selecting only pixels that have already cleared
a positive noise excursion, while discarding the symmetric negative excursions expected from the
same noise, systematically inflates the total relative to the true source flux. These two numbers
should therefore be read only as an upper bound on what the visible patches would represent if every
one of them were a real, coincident source. Second, and statistically more meaningful, is the net
signal integrated over all 30,096 pixels of the box without any positivity selection -- an unbiased
estimator of the true diffuse flux (or lack thereof) enclosed by the structure. Of these pixels,
12.0\% (3613, dominated by a single $\sim$3.6~arcmin$^2$ patch) are masked; this dominant patch is
centered within 12\arcsec\ of a foreground ($z=0.98$) CW-HCG group (undetected, SNR$=0.84$) rather
than overlapping any part of the AmicOne redshift range, so the masking does not preferentially
remove flux belonging to the structure under test. Propagating the per-pixel error map in quadrature
over the unmasked pixels gives a net count rate of $(2.23\pm4.96)\times10^{-4}$~counts~s$^{-1}$
(SNR$=0.45$), i.e.\ a flux of $(1.21\pm2.68)\times10^{-15}$~erg~cm$^{-2}$~s$^{-1}$ and $L_{\rm X} =
(2.0\pm4.4)\times10^{43}$~erg~s$^{-1}$ ($\log_{10}(L_{\rm X}/\mathrm{erg~s}^{-1}) = 43.30$), fully
consistent with zero, with a corresponding $3\sigma$ upper limit of
$\log_{10}(L_{\rm X}/\mathrm{erg~s}^{-1}) < 44.12$ at $z=2.65$ for the full structure. The
$\sim$0.2--1~dex gap between the (biased) patch-sum estimate and the (unbiased) whole-box $3\sigma$
upper limit itself demonstrates that the $2\sigma$ patches in Fig.~\ref{fig:amicone_null} are
dominated by positive noise fluctuations rather than tracing a single coherent, structure-scale
source, consistent with the independent tile-based null test below.

We then tiled the AmicOne box itself with an independent, non-overlapping grid of 300~kpc apertures
(100 tiles) on this same masked map, and separately built a field-calibrated null distribution by
repeating the identical tiling procedure across the full mosaic ($\sim$12,000 tiles), since the
per-pixel noise at this aperture scale is not simply Gaussian. Four of the 100 tiles reach
SNR~$\geq2.0$ and one reaches SNR~$\geq2.7$, fully consistent with the field-calibrated expectation
(2.9 and 0.9 tiles expected by chance out of 100, respectively; Poisson $p=0.33$ and $p=0.58$). The
single highest-significance tile (SNR~$=3.06$, RA$=150.2669\degr$, Dec$=1.9237\degr$,
$\log_{10}(L_{\rm X}/\mathrm{erg~s}^{-1})=43.8$) is not statistically significant once the number of
independent tiles searched is accounted for. Patches at the $2\sigma$ level are present throughout
the box, including near its northern, southern, and western edges (Fig.~\ref{fig:amicone_null}),
at a rate consistent with the field-calibrated null rather than with a single coherent,
structure-scale feature. We therefore do not detect a statistically significant diffuse signal from
the AmicOne structure at the current depth; given the strong $(1+z)^4$ cosmological
surface-brightness dimming expected at $z\sim2.5$--3, genuine intragroup or intergalactic gas
associated with this structure, if present, would plausibly be faint and spatially patchy rather
than smoothly extended, and we cannot rule out emission below our current per-tile sensitivity.
Deeper or stacked observations would be needed to test this directly.

As a further test that does not depend on AMICO group membership at this redshift, we do not rely on
the 15 CW-All group positions -- whose
group assignment at $z\sim2.5$--3 may itself be uncertain -- but instead map the projected galaxy
overdensity directly from the full COSMOS-Web photometric catalog (784,016 sources;
\citealt{Shuntov2025})
within the AmicOne box, selecting galaxies with \texttt{LP\_type}$=0$ and \texttt{MAG\_MODEL\_F444W}
$<27.5$. This overdensity map is itself an unoptimized diagnostic relative to a matched-filter
detector such as AMICO: we use a simple fixed-width Gaussian kernel on a $200\times200$ grid (pixel
scale $3.4\arcsec$, smoothing $\sigma = 3$~pixels $\approx10.3\arcsec$, uniform across the box) rather
than AMICO's own optimal, redshift- and density-adaptive filter, so the resulting $\delta$ map is
correspondingly noisier and not tuned to recover the specific clumps AMICO identifies; it is intended
only as an independent, coarse tracer of the projected galaxy density, not a competing detection
method. Because the AmicOne redshift window ($2.50\leq z<3.09$) is broad, we further test whether
any narrower sub-slice better isolates the structure by repeating the overdensity map in five sliding
$\Delta z=0.20$ slices (2.50--2.70, 2.60--2.80, 2.70--2.90, 2.80--3.00, and 2.90--3.10) as well as
over the full window. In each case we bin the selected galaxies onto this grid, Gaussian-
smooth the counts, and compute the fractional overdensity $\delta=(\Sigma-\bar\Sigma)/\bar\Sigma$;
we then regrid the source-masked X-ray significance map onto the same pixel grid and compute the
Pearson cross-correlation coefficient $r$ between $\delta$ and the X-ray significance. Pearson's $r$
tests only for a linear association between the two maps; given the essentially null X-ray signal
established above, this test has limited statistical power by construction, since a genuinely
informative version would require significant X-ray signal to correlate against in the first place.
What it can still do is check whether the noise-level $2\sigma$ patches in
Fig.~\ref{fig:amicone_null} happen to line up spatially with the independently-selected galaxy
overdensity peaks, which would be a coincidence worth flagging even in the absence of a formal
detection. Across all six
slices, $r$ ranges from $-0.07$ to $-0.01$ (Fig.~\ref{fig:amicone_null}): the galaxy
overdensity peaks and the X-ray significance peaks are not spatially coincident in any redshift
slice, and if anything the (very weak) association is slightly negative rather than positive. No
sub-slice recovers a preferred redshift at which the X-ray patches align better with the galaxy
density field. This is consistent with, but does not on its own strengthen, the whole-box null result
above: the X-ray map alone is not a strong diagnostic for a genuinely diffuse, low-surface-brightness
structure at this redshift, especially if the structure is not yet virialized or is spatially
fragmented into the 14 individual cores reported by \citet{Toni2025} rather than smoothly extended, and
we do not claim this cross-correlation test independently rules out diffuse emission -- only that it
finds no positional coincidence between the marginal $2\sigma$ X-ray patches and the galaxy density
field.

\section{Robustness of the $L^\star$ evolution}
\label{app:xlf_robustness}

Because each Schechter fit has three free parameters and, in several bins, fewer than 100
detections, we test whether the monotonic $\log L^\star$ rise reported in
Sect.~\ref{sec:xlf_schechter} is an artifact of parameter degeneracy or of which particular
groups happen to populate a bin, rather than a stable feature of the data. We repeat the fits in
two ways: (i) with $\alpha$ fixed to a single reference value ($\alpha = -1.0$) rather than free,
refitting only $\log\Phi^\star$ and $\log L^\star$ with a least-squares estimator, to check whether
the $\log\Phi^\star$--$\log L^\star$--$\alpha$ degeneracy visible in several corner plots (e.g.\
Fig.~\ref{fig:xlf_corner}) is driving the apparent trend; and (ii) with a bootstrap over the
detected group list itself (200 resamples with replacement per bin, refit with the same
fixed-$\alpha$ estimator for tractability), which tests sensitivity to sample composition rather
than to the fitting method.

Fixing $\alpha=-1.0$ reproduces the same monotonic rise seen in the free-$\alpha$ fits, at every
redshift and in both catalogs: $\log L^\star = 42.29, 43.30, 43.75, 44.65$ for CW-All and $42.52,
43.05, 44.28, 44.91$ for CW-HCG across the four bins, respectively, within $\lesssim$0.2~dex of the
free-$\alpha$ values in Table~\ref{tab:xlf_schechter} in every bin. The bootstrap gives consistent
medians (e.g.\ CW-All: $42.19\pm0.08$, $43.24\pm0.17$, $43.73\pm0.10$, $44.67\pm0.12$) with scatter
comparable to or smaller than the MCMC credible intervals in six of the eight bin/catalog
combinations, confirming the rise is not an artifact of the specific groups sampled. The exception
is the CW-HCG $2.20 \leq z < 3.80$ bin, where the bootstrap scatter ($\sigma = 1.25$~dex) is far
larger than its already-broad MCMC credible interval ($^{+0.27}_{-0.24}$~dex): with only 36
detections, this single bin's fit is materially sensitive to which handful of the most luminous
groups are resampled, and its $\log L^\star$ value should be treated as indicative rather than a
precision measurement. We flag this explicitly rather than omit the bin, since the qualitative rise
is still recovered in every resampling that converges; only its precise numerical value is unstable.

\begin{figure*}[htbp]
\centering
\includegraphics[width=0.98\textwidth]{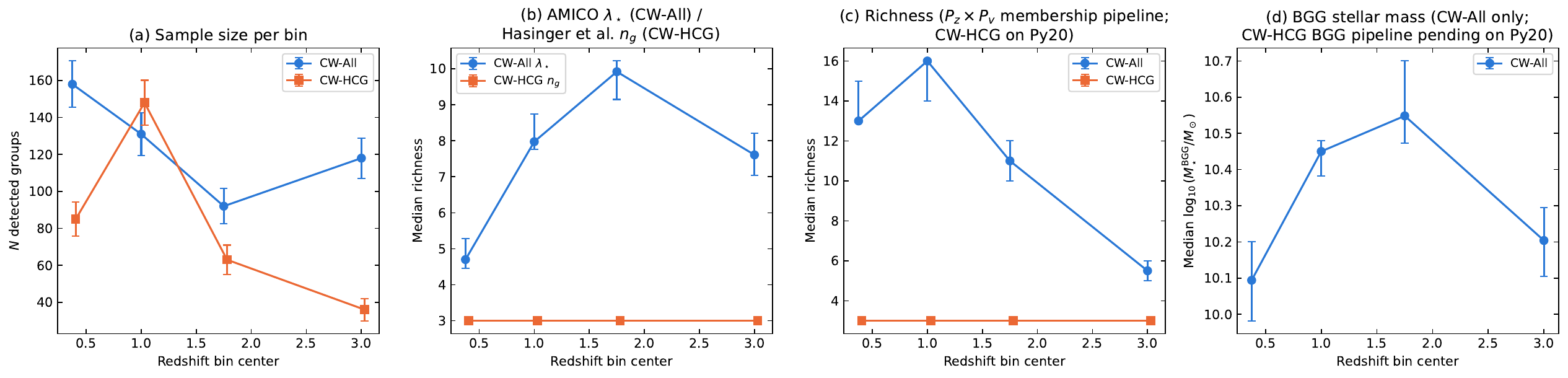}
\caption{Sample composition versus redshift bin, for direct comparison with the $L^\star$ evolution
of Table~\ref{tab:xlf_schechter}. (a) Number of X-ray-detected groups per bin; error bars are
Poisson ($\sqrt{N}$). (b) Median AMICO intrinsic richness $\lambda_\star$ for CW-All
\citep{Toni2025}, and, for CW-HCG, Hasinger et al.'s own discovery-time richness $n_g$ from the
parent compact-group catalog (Sect.~\ref{sec:groupcat}). (c) Median richness from our own
$P_z\times P_v$ membership pipeline (Sect.~\ref{sec:membership}, independent of $\lambda_\star$) for
CW-All, and $n_g$ again for CW-HCG. (d) Median BGG
stellar mass for CW-All (the equivalent CW-HCG measurement requires a BGG-selection pipeline run
not yet available for the current CW-HCG catalog). Error bars in (b)--(d) are 68\% bootstrap
confidence intervals on the median itself (2000 resamples), not the population scatter. CW-All's
$\lambda_\star$ is flat within the bin-to-bin scatter with no monotonic trend, and its membership
richness \emph{declines} with redshift, while $L^\star$ rises; CW-HCG's discovery richness $n_g$ is
essentially flat at its selection floor ($n_g=3$ in every bin) throughout; CW-All's BGG mass is
also flat. None of the available proxies tracks the $L^\star$ evolution in the direction that
mass/richness selection would predict, supporting the conclusion that halo-mass/richness selection
alone does not explain the observed $L^\star$ evolution.}
\label{fig:xlf_sample_comp}
\end{figure*}

To determine whether the rise between individual bins is statistically distinguishable from no
evolution, we compute the full difference distribution $\Delta\log L^\star = \log L^\star_{j} - \log
L^\star_{i}$ between each pair of adjacent bins directly from their free-$\alpha$ MCMC posteriors
(300,000 random posterior-sample pairs per comparison), rather than comparing point estimates alone;
this propagates the complete, possibly skewed, shape of each bin's posterior into the significance
of the difference. Table~\ref{tab:xlf_significance} reports the resulting median $\Delta\log
L^\star$, its 68\% and 95\% credible intervals, and the posterior probability $P(\Delta \leq 0)$
that the higher-redshift bin is \emph{not} more luminous than the lower-redshift bin. The evidence
for evolution is uneven across the four bins rather than uniformly strong: the CW-All $1.30$--$2.20
\to 2.20$--$3.80$ step and the CW-HCG $0.70$--$1.30 \to 1.30$--$2.20$ step are each significant well
beyond 95\% confidence ($P(\Delta\leq0) = 0.0025$ and $0.0036$, respectively), and can be considered
robust detections of evolution. The lowest-redshift step ($0.05$--$0.70 \to 0.70$--$1.30$) is only
marginal in both catalogs (95\% credible intervals include zero), driven by the broad, right-skewed
posterior of the $0.05\leq z<0.70$ bin itself (Table~\ref{tab:xlf_schechter}). Comparing the lowest
and highest redshift bins directly (the full baseline spanned by our four-bin scheme) gives
$\Delta\log L^\star = 2.15$~dex for CW-All (95\% CI $[0.03, 2.61]$, $P(\Delta\leq0)=0.024$) and
$2.25$~dex for CW-HCG (95\% CI $[-0.21, 2.98]$, $P(\Delta\leq0)=0.043$): both are consistent with
net positive evolution over the full baseline at close to the 95\% level, but the credible intervals
extend close to, and in the CW-HCG case fractionally across, zero -- a materially weaker statement
than the $\sim$2~dex point-estimate difference alone would suggest. We therefore conclude that the
data support genuine $L^\star$ evolution with high confidence in the upper half of our redshift
range, and are suggestive of, but do not tightly constrain, evolution in the lower half; this uneven
pattern is a direct consequence of bin-to-bin differences in sample size and posterior shape rather
than of the fitting method (Table~\ref{tab:xlf_schechter}).

\begin{table}[!htb]
\caption{Posterior-based significance of the $\log L^\star$ rise between adjacent redshift bins (free-$\alpha$ MCMC
  posteriors, 300,000 paired draws per row). $\Delta\log L^\star$ is the median of the difference distribution between the
  higher- and lower-redshift bin in each row; credible intervals are on this difference. $P(\Delta\leq0)$ is the posterior
  probability that the higher-redshift bin is not more luminous.}
\label{tab:xlf_significance}
\centering
\footnotesize
\setlength{\tabcolsep}{3pt}
\begin{tabular}{llccc}
\hline\hline
Catalog & Comparison & $\Delta\log L^\star$ & 95\% CI & $P(\Delta\leq0)$ \\
\hline
CW-All & $0.05$--$0.70 \to 0.70$--$1.30$ & $+0.78$ & $[-1.34, 1.19]$ & $0.081$ \\
CW-All & $0.70$--$1.30 \to 1.30$--$2.20$ & $+0.27$ & $[-0.08, 0.68]$ & $0.056$ \\
CW-All & $1.30$--$2.20 \to 2.20$--$3.80$ & $+1.10$ & $[0.68, 1.50]$ & $0.0025$ \\
CW-All & $0.05$--$0.70 \to 2.20$--$3.80$ (full) & $+2.15$ & $[0.03, 2.61]$ & $0.024$ \\
\hline
CW-HCG & $0.05$--$0.70 \to 0.70$--$1.30$ & $+0.22$ & $[-2.26, 0.75]$ & $0.354$ \\
CW-HCG & $0.70$--$1.30 \to 1.30$--$2.20$ & $+0.57$ & $[0.17, 2.23]$ & $0.0036$ \\
CW-HCG & $1.30$--$2.20 \to 2.20$--$3.80$ & $+1.46$ & $[-0.17, 2.06]$ & $0.039$ \\
CW-HCG & $0.05$--$0.70 \to 2.20$--$3.80$ (full) & $+2.25$ & $[-0.21, 2.98]$ & $0.043$ \\
\hline
\end{tabular}
\end{table}

\end{appendix}

\end{document}